
%
%

\input harvmac
\input pictex
\def\sm{$S$-matrix}\def\sms{$S$-matrices}
\def\scth{scattering theory}\def\scths{scattering theories}

\def\smt{$S$-matrix theory}
\def\smes{$S$-matrix elements}
\def\tm{transfer matrix}\def\tms{transfer matrices}
\def\BA{Bethe Ansatz}\def\BY{Bethe-Yang}
\def\Zam{Zamolodchikov}
\def\ie{{\it i.e.}}
\def\eg{{\it e.g.}}
\def\at{\tilde{a}}\def\bt{\tilde{b}}\def\ct{\tilde{c}}
\def\dt{\tilde{d}}
\def\At{\tilde{A}}\def\Bt{\tilde{B}}\def\Ct{\tilde{C}}\def\Dt{\tilde{D}}
\def\no{\noindent}
\def\o{\over}
\def\ra{\rangle}
\def\sc{\scriptstyle}
\def\scsc{\scriptscriptstyle}

\def\nl{\hfill\break}
\def\alp{\alpha}\def\gam{\gamma}\def\del{\delta}\def\lam{\lambda}
\def\eps{\epsilon}\def\veps{\varepsilon}
\def\sig{\sigma}\def\th{\theta}
\def\thb{{\th\kern-0.465em \th}}

\def\ZZ{{\bf Z}}
\def\NN{{\bf N}}
\def\RR{{\bf R}}

\def\ontopss#1#2#3#4{\raise#4ex \hbox{#1}\mkern-#3mu {#2}}

\setbox\strutbox=\hbox{\vrule height12pt depth5pt width0pt}
\def\tablerule{\noalign{\hrule}}
\def\strut{\relax\ifmmode\copy\strutbox\else\unhcopy\strutbox\fi}
%

\nref\rZamkink{A.B.~Zamolodchikov, Landau Institute preprint (1989)}
\nref\rRSG{A.~LeClair, Phys.~Lett.~230B (1989) 103;\nl
  D.~Bernard and A.~LeClair, Nucl.~Phys.~B340 (1990) 721;\nl
  N.Yu.~Reshetikhin and F.A.~Smirnov, Comm.~Math.~Phys.~131 (1990) 157}
\nref\rSUSYi{K.~Schoutens, Nucl.~Phys.~B344 (1990) 665;
 C.~Ahn, Nucl.~Phys. B354 (1991) 57}
\nref\rSUSYii{P.~Fendley,
 S.D.~Mathur, C.~Vafa and N.P.~Warner, Phys.~Lett.~243B (1990) 257}
\nref\rSmir{F.A.~Smirnov, Int.~J.~Mod.~Phys.~A6 (1991) 1407}
\nref\rslmp{A.B.~Zamolodchikov, ``\sm\ of the Subleading Magnetic
 Perturbation Of the Tricritical Ising Model'', Princeton preprint
 PUPT 1195 (1990)}
\nref\rTanig{A.B.~Zamolodchikov, Adv.~Stud.~Pure Math.~19 (1989) 1}
\nref\rYZ{V.P.~Yurov and Al.B.~Zamolodchikov, Int.~J.~Mod.~Phys.~A5 (1990)
 3221, and Paris preprint ENS-LPS-273 (1990)}
\nref\rtunnel{V.~Privman and M.E.~Fisher, J.~Stat.~Phys.~33 (1983) 285; \nl
 E.~Br\'ezin and J.~Zinn-Justin, Nucl.~Phys.~B257 (1985) 867; \nl
 G.~M\"unster, Nucl.~Phys.~B324 (1989) 630}
\nref\rLui{M.~L\"uscher, in: {\it Progress in Gauge Field Theory}
 (Cargese 1983), ed. G.~'t~Hooft et al (Plenum, New York, 1984), and
 Commun.~Math.~Phys.~104 (1986) 177}
\nref\rouriv{T.R.~Klassen and E.~Melzer, Nucl.~Phys.~B362 (1991) 329}
\nref\rYangii{C.N.~Yang and C.P.~Yang, J.~Math.~Phys.~10 (1969) 1115}
\nref\rZamtba{Al.B.~Zamolodchikov, Nucl.~Phys.~B342 (1990) 695}
\nref\rourii{T.R.~Klassen and E.~Melzer, Nucl.~Phys.~B338 (1990) 485}
\nref\rouriii{T.R.~Klassen and E.~Melzer, Nucl.~Phys.~B350 (1991) 635}
\nref\rMartexc{M.J.~Martins, Phys.~Lett.~257B (1991) 317, and
  Phys.~Rev.~Lett.~67 (1991) 419}
\nref\rourv{T.R.~Klassen and E.~Melzer, ``Spectral Flow between
  Conformal Field Theories in 1+1 Dimensions'', Chicago/Miami preprint
  EFI 91-17/UMTG-162 (1991), Nucl.~Phys.~B (in  press)}
\nref\rFend{P.~Fendley, ``Excited-state Thermodynamics'', Boston preprint
  BUHEP-91-16 (1991)}
\nref\rRSOStba{Al.B.~Zamolodchikov, Nucl.~Phys.~B358 (1991) 497}
\nref\rFenInt{P.~Fendley and K.~Intriligator,
 ``Scattering and Thermodynamics of Fractionally-Charged Supersymmetric
 Solitons'', Boston/Harvard preprint
 BUHEP-91-17/HUTP-91-A043 (1991), ~``Scattering and Thermodynamics in
 Integrable $N=2$ Theories'', BUHEP-92-5/HUTP-91-A067 (1992)}
\nref\rLuii{M.~L\"uscher, Commun.~Math.~Phys.~105 (1986) 153, and
  Nucl.~Phys.~B354 (1991) 531}
\nref\rLasMar{M.~L\"assig and M.J.~Martins, Nucl.~Phys.~B354 (1991)
 666;\nl M.J.~Martins, Phys.~Lett.~262B (1991) 39}
\nref\rLuWo{M.~L\"uscher and U.~Wolff, Nucl.~Phys.~B339 (1990) 222}
\nref\rZams{A.B.~Zamolodchikov and Al.B.~Zamolodchikov, Ann.~Phys.~120
 (1980) 253}
\nref\rKar{M.~Karowski, Nucl.~Phys.~B153 (1979) 244}
\nref\rSTW{B.~Schroer, T.T.~Truong and P.~Weisz, Phys.~Lett.~63B (1976) 422}
\nref\rYang{C.N.~Yang, Phys.~Rev.~Lett.~19 (1967) 1312}
\nref\rBax{R.J.~Baxter, {\it Exactly Solved Models in Statistical mechanics}
 (Academic Press, London, 1982);
 R.J.~Baxter and P.A.~Pearce, J.~Phys.~A15 (1982) 897}
\nref\rBPZ{A.A.~Belavin, A.M.~Polyakov and A.B.~Zamolodchikov,
 Nucl.~Phys.~B241 (1984) 333; \nl
 D.~Friedan, Z.~Qiu and S.H.~Shenker, Phys.~Rev.~Lett.~52 (1984)
 1575; \nl  P.~Goddard, A.~Kent and D.~Olive, Phys.~Lett.~152B (1985) 88}
\nref\rModInv{J.L.~Cardy, Nucl.~Phys.~B270 (1986) 186;
 {\it ibid.}~B275 (1986) 200; \nl
 A.~Cappelli, C.~Itzykson and J.-B.~Zuber, Nucl.~Phys.~B280 (1987) 445; \nl
 D.~Gepner, Nucl.~Phys.~B287 (1987) 111}
\nref\rZamLG{A.B.~Zamolodchikov, Sov.~J.~Nucl.~Phys.~44 (1986) 529}
\nref\rLudCar{A.W.W.~Ludwig and J.L.~Cardy, Nucl.~Phys.~B285 (1987) 687}
\nref\rKMS{D.A.~Kastor, E.J.~Martinec and S.H.~Shenker,
 Nucl.~Phys.~B316 (1989) 590}
\nref\rNeqtwo{E.J.~Martinec, Phys.~Lett.~217B (1989) 431;\nl  C.~Vafa and
 N.P.~Warner, Phys.~Lett.~218B (1989) 51}
\nref\rEguYan{T.~Eguchi and S.-K.~Yang, Phys.~Lett.~235B (1990) 282}
\nref\rAGS{L.~Alvarez-Gaum\'e, C.~Gomez and G.~Sierra,
           Phys.~Lett.~220B (1989) 142}
\nref\rIFT{T.T.~Wu, B.M.~McCoy, C.A.~Tracy and E.~Baruch, Phys.~Rev.~B13
 (1976) 316;\nl  M.~Sato, T.~Miwa and M.~Jimbo, Proc.~Japan Acad.~53A
 (1977) 6, 147, 153, 183,  219; \nl
 B.~Schroer and T.T.~Truong, Nucl.~Phys.~B144 (1978) 80}
\nref\rZamflow{A.B.~Zamolodchikov, Sov.~J.~Nucl.~Phys.~46 (1987) 1090}
\nref\rmlesstba{Al.B.~Zamolodchikov, Nucl.~Phys.~B358 (1991) 524}
\nref\rRav{F.~Ravanini, ``RG Flows of Non-diagonal Minimal Models Perturbed
  by $\phi_{1,3}$'', Paris preprint SPhT/91-147 (1991)}
\nref\rourvii{T.R.~Klassen and E.~Melzer, ``RG Flows in the $D$-Series of
 Minimal CFTs'', Cornell/Stony Brook preprint CLNS-91-1111/ITP-SB-91-57
 (1991)}
\nref\rCarLH88{J.L.~Cardy, in: {\it Fields, Strings, and Critical Phenomena},
 Les Houches 1988,
 ed. E.~Br\'ezin and J.~Zinn-Justin, (North Holland, Amsterdam, 1989)}
\nref\rorbif{P.~Fendley and P.~Ginsparg, Nucl.~Phys.~B324 (1989) 549}
\nref\rPasq{V.~Pasquier, Nucl.~Phys.~B285 (1987) 162}
\nref\rFerFi{A.E.~Ferdinand and M.E.~Fisher, Phys.~Rev.~185 (1969) 832;\nl
 H.~Saleur and C.~Itzykson, J.~Stat.~Phys.~48 (1987) 449}
\nref\rCardy{J.L.~Cardy, J.~Phys.~A17~(1984)~L385}
\nref\rRocha{A.~Rocha-Caridi, in: {\it Vertex Operators in Mathematics and
 Physics}, ed. J.~Lepowsky et al (Springer, New York, 1985)}
\nref\rLMC{M.~L\"assig, G.~Mussardo and J.L.~Cardy, Nucl.~Phys.~B348 (1991)
 591}
\nref\rLasMus{M.~L\"assig and G.~Mussardo, Computer Phys.~Comm.~66 (1991)
 71}
\nref\rMuss{F.~Colomo, A.~Koubek and G.~Mussardo, ``On the \sm\ of the
 Subleading Magnetic Deformation of the Tricritical Ising Model in Two
 Dimensions'', preprint ISAS/94/91/EP (1991)}
\nref\rKobSwi{R.~K\"oberle and J.A.~Swieca, Phys.~Lett.~86B (1979) 209}
\nref\rIJMP{A.B.~Zamolodchikov, Int.~J.~Mod.~Phys.~A3 (1988) 743}
\nref\rArin{A.E.~Arinshtein, V.A.~Fateev and A.B.~Zamolodchikov,
  Phys.~Lett.~87B (1979) 389}
\nref\rCMouri{J.L.~Cardy and G.~Mussardo, Phys.~Lett.~225B (1989) 275; \nl
 P.G.O.~Freund, T.R.~Klassen and E.~Melzer, Phys.~Lett.~229B (1989) 243}
\nref\rFendN{P.~Fendley, in preperation}
\nref\rSwi{J.A.~Swieca, Phys.~Rev.~D13 (1976) 312}
\nref\rGinsp{P.~Ginsparg, Nucl.~Phys.~B295 (1988) 153}

\Title{\vbox{\baselineskip12pt\hbox{CLNS-92/1130~~~ITP-SB-92-01}
\hbox{hepth@xxx/9202034}}}
{\vbox{\centerline{Kinks in Finite Volume}}}
\centerline{Timothy R.~Klassen\foot{email: klassen@strange.tn.cornell.edu}}
\smallskip\centerline{\it Newman Laboratory for Nuclear Studies}
\smallskip\centerline{\it Cornell University, Ithaca, NY 14853}
\medskip\centerline{and}
\medskip\centerline{Ezer Melzer\foot{email: melzer@max.physics.sunysb.edu}}
\smallskip\centerline{\it Institute for Theoretical Physics}
\smallskip\centerline{\it SUNY, Stony Brook,  NY 11794-3840}
\vskip 5mm

A (1+1)-dimensional quantum field theory with a degenerate vacuum
(in infinite volume) can contain particles, known as kinks, which
interpolate between different vacua and have nontrivial restrictions on
their multi-particle Hilbert space. Assuming such a theory
to be integrable, we show how to calculate the multi-kink energy levels
in finite volume given its factorizable $S$-matrix.
In massive theories this can be done exactly up to contributions due to
off-shell and tunneling effects that fall off exponentially with volume.
As a first application we compare our analytical predictions for the kink
scattering theories conjectured to describe the subleading thermal and
magnetic perturbations of the tricritical Ising model with numerical
results from the truncated conformal space approach.
In particular, for the subleading magnetic perturbation
our results allow us to decide between the two
different $S$-matrices proposed  by Smirnov and Zamolodchikov.

\Date{\hfill 2/92}
\vfill\eject

\newsec{Introduction}
\ftno=0

A unique property of 2-dimensional Minkowski space is that a double light cone
divides the set of all space-like points into two disconnected components in
a Lorentz-invariant way. In a (1+1)-dimensional quantum field theory (QFT)
with degenerate vacuum it is therefore possible that the expectation value
of a field approaches different values at the two space-like infinities.
If the corresponding excitation creates a stable particle in the theory it is
generically known as a ``kink''; it is labelled by the two (possibly identical)
vacua between which it interpolates, in addition to its energy-momentum and
perhaps other quantum numbers.

Some novel and interesting features arise when one considers the multi-particle
Hilbert space of a theory with kinks. Since a multi-kink state corresponds to
a sequence of vacua on a line, one cannot arbitrarily compose single kinks
to obtain an allowed state. In other words, there are restrictions on the
multi-kink Hilbert space --- it is {\it not} a (direct sum of) bosonic or
fermionic Fock spaces.

Note that the solitons of the sine-Gordon model are not quite examples of
what we want to call a kink.
In this theory field configurations differing by
a period of the cosine potential are identified,
so that there is only one type of soliton and anti-soliton with {\it no}
restrictions (except the exclusion principle) on their multi-particle states.
(See however sect.~5 for a discussion of some variants of the
sine-Gordon model where such restrictions do arise.)
We reserve the term kink for
excitations with {\it nontrivial} restrictions on their multi-particle
states. Recently several~\rZamkink\rRSG\rSUSYi\rSUSYii\rSmir\rslmp\
factorizable \scths\ of massive
kinks have been proposed to describe certain integrable perturbations~\rTanig\
of conformal field theories (CFTs).

The purpose of this paper is to analyze the multi-kink spectrum of such
theories in finite volume with periodic boundary conditions, \ie~on a
circle of circumference $R$, say. We will show how to calculate
the energy levels exactly --- up to terms
due to tunneling and off-shell effects that are exponentially small for
large $R$ --- in terms of the factorizable \sm\ of the theory.

A theoretical understanding of the finite-volume spectrum of a QFT provides
a handle on the dynamics of a theory, allowing one, for instance,
to test a conjectured exact \sm.
The idea is to compare the analytical large-volume
predictions obtained from the \sm\ with one of the available numerical
methods of calculating the small-volume spectrum of a QFT, \eg~lattice
simulations or the ``truncated conformal space approach''~\rYZ\ (TCSA) for
a perturbed CFT (cf.~subsect.~3.3).

There are other methods of checking a conjectured \sm, involving comparison
with the finite-volume energy levels of 0- and 1-particle states.
For a generic theory of kinks these tests are not so easy to perform, and
here the multi-particle spectrum can provide a very simple way
of checking a proposed \sm.  In most of the remainder of this
introduction we will briefly review the basic physical phenomena
underlying the finite-volume spectrum of a QFT, and
what information about the \sm\ of the theory can be
gained by analyzing different parts of the spectrum.

\medskip
There are basically three physical effects contributing to the finite-volume
energy levels of a QFT, which we will refer to as ~(i) tunneling, ~(ii)
off-shell, and ~(iii) boundary and scattering effects.

\medskip
\no {\bf (i):} Tunneling, well-known from
quantum mechanics, occurs in theories with a
degenerate vacuum in infinite volume. Since the `energy barrier' between the
vacua is proportional to the volume of the world, there will be instantons
in finite volume, tunneling between
the vacua and lifting their exact degeneracy. In simple situations, like
$\phi^4$ theory in $d+1$ dimensions with `space' a hypercube of volume $R^d$,
the splitting (in the phase of spontaneously broken symmetry)
is~\rtunnel\ of ${\cal O}(R^{-1} R^{d/2} e^{-\sig R^d})$.
Here
the exponential is simply the Boltzmann factor of the instanton action,
the power $R^{d/2}$ is due to zero modes, and $R^{-1}$ is due to 1-loop
fluctuations. In 1+1 dimensions the exponential is $e^{-m R}$, $m$ being
the mass of the kink. The prefactor in the above formula can be calculated
perturbatively (see the last ref.~in~\rtunnel) in a given QFT, but it is
apparently not known how to express it in a ``universal'' manner in terms
of the \sm\ of a generic theory --- in fact, it is not clear if such a
universal formula exists.  If there is tunneling, it is not only relevant
for the energy levels of 0-particle states; its effects on other levels will
be of comparable magnitude, although we are not aware of any explicit
calculations.

\medskip
\no {\bf (ii):}
There are two ways in which off-shell effects influence
the finite-volume spectrum.
The first, relevant for multi-particle states, reflects the fact
that in finite volume interactions between  physical particles (due to the
exchange of virtual particles) cannot be completely expressed in terms of
\smes, which connect {\it asymptotic} states.
 Since in a massive theory interactions decrease exponentially with
the distance between the particles,
the contributions of this effect to the energy levels of stationary scattering
states (cf.~(iii) and subsect.~2.2) are expected to
decrease exponentially with volume.
The second off-shell effect is due to the fact that in finite volume
virtual particles can ``travel once or several
times around the world'' before annihilating again
or being absorbed by a real particle.
The resulting volume dependence of these
{\it vacuum polarization} effects again
gives rise to exponentially small corrections to the energy levels.
For the case of
zero momentum 1-particle energies, \ie~finite-volume masses, these
corrections have been studied in some detail~\rLui\rouriv\ for essentially
arbitrary massive QFTs
(not only    integrable theories in 1+1 dimensions)
without a degenerate vacuum in infinite volume.
The leading terms contributing to finite-size mass corrections in such theories
are given by a universal formula involving only \smes.
It is not known if the {\it exact} finite-size mass corrections can
also be expressed solely through scattering amplitudes.

For 0-particle states stronger results can be obtained in certain cases.
In 1+1 dimensions there is a beautiful method, known as the thermodynamic
Bethe Ansatz~\rYangii\rZamtba\rourii\rouriii\ (TBA), to calculate the
{\it exact} finite-volume
ground state energy $E_0(R)$ of an integrable QFT in terms of its \sm.
(And with some modifications this method can also be applied to certain
0-particle states above the ground state~\rMartexc\rourv\rFend.)
Although restricted to integrable theories, this method does not require the
existence of a unique vacuum. The exact calculation of the ground state
energy on a circle is possible due to its relation
to the free energy on an infinite
line at finite temperature (and zero chemical potentials).
The final result for $E_0(R)$, expressed in terms of the solution of a
set of coupled non-linear integral equations, can provide a
very strong test of
a conjectured \sm, since the small-$R$ behaviour of $E_0(R)$ allows
one to extract the central charge and other properties of the CFT describing
the UV limit of the QFT under consideration.\foot{However, in some cases
$E_0(R)$ does not uniquely determine the UV CFT; this happens
\eg~for certain perturbations of  minimal CFTs
with the same central charge but different
modular invariant partition function.}
Unfortunately, if the \sm\ of the theory is not diagonal,
the TBA requires a
detailed case by case analysis (see~\rRSOStba\rFenInt\ for the first
examples). Such an analysis
can be quite nontrivial for kink theories,  and
therefore other methods of checking a conjectured \sm\ are desirable.

\medskip
\no {\bf (iii):}
   From quantum mechanics one is familiar with the fact that
boundary conditions lead to the quantization of energies and momenta.
For 1-particle states in a QFT       this effect,
together with contributions due to tunneling and vacuum polarization,
completely determines the energies in finite volume. For multi-particle states
one must also take scattering effects into account, which simply
refer to the fact that energy eigenstates
in finite volume are stationary scattering states.\foot{In
integrable theories states with a definite number of particles will not
mix with states of different particle number,
whereas for non-integrable theories this will
be true generically (\ie~in the absence of appropriate selection rules) only
for 2-particle states with energies below the 3-particle threshold.
See~\rLuii\ for a detailed study of 2-particle levels in higher-dimensional
(and therefore non-integrable) theories.}
The resulting quantization conditions, which we will refer to as
{\it Bethe Ansatz equations}, can be expressed solely
in terms of the \sm\ of the (factorizable) theory
if we neglect the nonzero range of the interactions.
The corresponding finite-size corrections to
energies are generically of ${\cal O}(1/R^2)$, as we will
see, and therefore much larger
than the exponentially small (in massive theories, at least)
contributions due to the finite interaction range,
off-shell, or tunneling effects.
Multi-particle energy levels are therefore ideally suited to provide a check
on kink \sms.

Previously multi-particle states
were only studied for theories with diagonal \sms~\rYZ\rLasMar; 2-particle
levels were also
discussed for the $O(3)$ non-linear $\sigma$-model~\rLuWo, where the \sm\
is diagonal in a 2-particle basis of isospin eigenstates.

\bigskip

The paper is organized as follows. In sect.~2 we
describe how to obtain  the energy levels and eigenfunctions of multi-kink
states in finite volume (up to
the exponentially small corrections discussed above).
Sects.~3 and 4 are devoted to examples.
In sect.~3 we start with a description of some general features of
$\phi_{1,3}$-perturbed unitary minimal CFTs, leading
(in one direction) to integrable theories of massive kinks.
As a warm-up, we discuss in subsect.~3.1
the simplest member of this family, the Ising field theory in
the phase of spontaneously broken symmetry.
For this theory the {\it complete} finite-volume spectrum is known
{\it exactly}, and provides an instructive  check  on the large-volume
Bethe Ansatz predictions
(which are rather trivial to obtain in this case).
 Subsect.~3.2  treats the next theory in the family,
the subleading thermal perturbation of the tricritical Ising CFT.
In subsect.~3.3 we compare our analytical predictions
for 2- and 4-kink states
with numerical results of the truncated conformal space approach (TCSA).
Sect.~4 deals with another family of
integrable theories of kinks, the $\phi_{2,1}$-perturbed unitary minimal
CFTs. We focus our attention on the subleading magnetic
perturbation of the tricritical Ising CFT,
for which two different \sm\ theories have been proposed, one by
Smirnov~\rSmir\ and the other by  Zamolodchikov~\rslmp.
We clarify the relation between the two $S$-matrices
and point out some {\it a priori} problems with Zamolodchikov's conjecture.
Comparing our large-volume results for 2- and 3-kink energy levels  with
TCSA data provides very strong support for Smirnov's proposal
 (although the subtle issue of ``crossing
factors'', cf.~subsects.~2.3 and~4.1, remains open).
In sect.~5 we present our conclusions, and discuss some simple modifications
of the sine-Gordon model which lead to theories with kinks.

\newsec{The $N$-Particle Transfer Matrix and the Bethe-Yang Equations}

\subsec{$S$-matrix conventions}

In an integrable massive
$(1+1)$-dimensional QFT all scattering amplitudes factorize into 2-particle
amplitudes ~$S_{ab}^{cd}(\th=|\th_{12}|)$~ describing scattering processes
{}~$a(\th_1)+b(\th_2) \to c(\th_2) + d(\th_1)$~ among the different
particles $a,b,\ldots$ of the theory
(see fig.~1(a)).
\vskip 7mm
\midinsert
\centerline{
\beginpicture
\setcoordinatesystem units <1pt,1pt>
\linethickness 4.0pt
\arrow <4pt> [.3,.5] from -80 -10 to -60 -10
\arrow <4pt> [.3,.5] from -80 -10 to -80  10
\put {$x$} at -53 -10
\put {$t$} at -80 16
\setlinear
\plot 0 20  40 -20 /
\plot 0 -20 40 20 /
\plot 120 20 160 -20 /
\plot 120 -20 160 20 /
\put {$d$} at 45 25
\put {$b$} at 45 -25
\put {$a$} at -5 -25
\put {$c$} at -5  25
\put {$\th_1$} at 6  -5
\put {$\th_2$} at 34 -5
\put {(a)} at 20 -45
\put {$\alp$} at 126 0
\put {$\beta$} at 155 0
\put {$\del$} at 140 14
\put {$\gam$} at 140 -15
\put {$\th_1$} at 117 -25
\put {$\th_2$} at 165 -25
\put {(b)} at 140 -45
\endpicture
}
\vskip 0.5cm
\no
\hbox{ {\bf Fig.~1:}
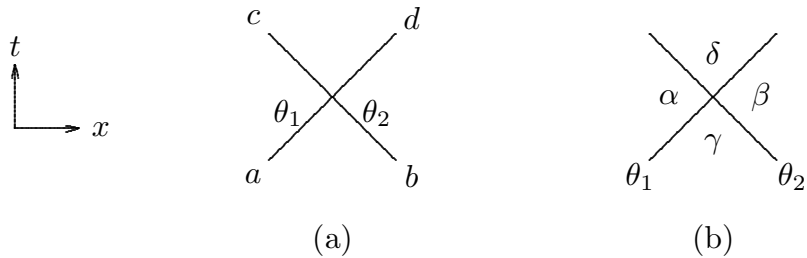
\vtop{\hbox{{\tenpoint (a) The 2-particle scattering process
 ~$a(\th_1)+b(\th_2)~ \to ~c(\th_2) + d(\th_1)$}}
\hbox{{\tenpoint (b) The 2-kink process
    ~$K_{\alpha\gam}(\th_1) + K_{\gam\beta}(\th_2) ~\to~
    K_{\alpha\del}(\th_2) + K_{\del\beta}(\th_1)$}} } }
\vskip 0.3cm
\endinsert

Here the $\th_i$ are rapidities, parametrizing the energy-momentum of a
particle on its mass shell via
\eqn\rap{ (p^0,p^1) ~=~ (m \cosh\th, ~m \sinh\th) ~, }
and ~$\th_{ij} \equiv \th_i -\th_j$.

We will be particularly interested in scattering theories of kinks. For such
theories it is more convenient to label the \smes\ by the asymptotic vacua
between which the kinks interpolate.
(To be sure, by essentially just rewriting
our results given below in kink language into ``ordinary'' \sm\
notation,  they apply to arbitrary factorizable \scths.)
We will consider the case of a single multiplet of kinks of mass $m>0$
which are all bosons or all fermions,
in order to keep the notation simple and
transparent; this assumption can be trivially
relaxed.
The kink interpolating between the vacuum $\alpha$
at $x\to -\infty$ and $\beta$ at $x\to +\infty$ will be denoted by
$K_{\alpha\beta}(\th)$.\foot{We will always use greek indices for the
asymptotic vacua;  this allows one to distinguish ``ordinary'' \smes, which
will be adorned by latin indices for the particles, from those in kink
notation.}
We let $S_{\alpha\beta}^{\gam\del}(\th_{12})$
describe the process
  $K_{\alpha\gam}(\th_1) + K_{\gam\beta}(\th_2) \to
   K_{\alpha\del}(\th_2) + K_{\del\beta}(\th_1)$,
where $\th_1 > \th_2$. The latter restriction ensures
the consistent ordering $\alp,\gam,\beta$ ($\alp,\del,\beta$) of the
vacua along space in the far past (future), as is clear from the
graphical description of the scattering process
presented in fig.~1(b).
(It will be convenient below to let $\alp,\beta,\gam,\del$ range over all
vacua and set $S_{\alp\beta}^{\gam\del}(\th)\equiv 0$ if one or more of the
kinks in fig.~1(b) does not exist in the theory.)

To get used to the kink notation it is instructive to rewrite the standard
requirements of \smt\ in this language. We will assume our theories to be
time-reversal and parity invariant, \ie\
$S_{\alpha\beta}^{\gam\del}(\th) = S_{\alpha\beta}^{\del\gam}(\th)
= S_{\beta\alpha}^{\gam\del}(\th)$.
Using {\it real analyticity} ~${S_{\alpha\beta}^{\gam\del}(\th)=
(S_{\alpha\beta}^{\gam\del}(-\th^\ast))^\ast}$,~
{\it unitarity} can        then      be written as
\eqn\unit{ \sum_{\mu} S_{\alpha\beta}^{\gam\mu}(\th)
    S_{\alpha\beta}^{\mu\del}(-\th) ~=~ \del^{\gam\del}
          ~~~~~~ {\rm for~all}~~\alp,\beta,\gam, \del ~.}
The requirement of
{\it crossing symmetry} amounts to
\eqn\crossing{S_{\alpha\beta}^{\gam\del}(\th) ~=~
              S_{\gam\del}^{\alpha\beta}(i\pi - \th) ~~.}
The {\it factorization} or {\it Yang-Baxter} equations~\rZams,
expressing the consistent factorization
of an arbitrary amplitude into 2-particle amplitudes, read
\eqn\YB{
 \sum_{\mu} S_{\alp\gam}^{\beta\mu}(\th_{12}) ~S_{\mu\del}^{\gam\eps}(\th_{13})
       ~S_{\alpha\eps}^{\mu\eta}(\th_{23}) ~=~
   \sum_{\mu} S_{\beta\del}^{\gam\mu}(\th_{23})~
          S^{\beta\eta}_{\alp\mu}(\th_{13})
         ~S_{\eta\del}^{\mu\eps}(\th_{12}) ~,}
which is illustrated in fig.~2.

\midinsert
\vskip 0.5cm
\centerline{
\beginpicture
\setcoordinatesystem units <1pt,1pt>
\linethickness 4.0pt
\arrow <4pt> [.3,.5] from -80 -10 to -60 -10
\arrow <4pt> [.3,.5] from -80 -10 to -80  10
\put {$x$} at -53 -10
\put {$t$} at -80 16
\setlinear
\plot 0 30  100 -20 /
\plot 33 -30  37 30 /
\plot 0 -30 100 20 /
\plot 140 20 240 -30 /
\plot 203 -30 207 30 /
\plot 140 -20 240 30 /
\put {$=$} at 120 0
\put {$\alp$} at 13 0
\put {$\beta$} at 26 -25
\put {$\eta$} at 28 25
\put {$\mu$} at 43  0
\put {$\gam$} at 60  -20
\put {$\eps$} at 62 20
\put {$\del$} at 88 0
\put {$\th_1$} at -3 -35
\put {$\th_2$} at 34 -37
\put {$\th_3$} at 106 -25
\put {$\alp$} at 152 0
\put {$\beta$} at 178 -20
\put {$\eta$} at 180 20
\put {$\mu$} at 196  0
\put {$\gam$} at 211  -25
\put {$\eps$} at 215 25
\put {$\del$} at 225 0
\put {$\th_1$} at 137 -25
\put {$\th_2$} at 204 -37
\put {$\th_3$} at 247 -35
\endpicture
}
\vskip 0.5cm
\no {\bf Fig.~2:}
{\tenpoint
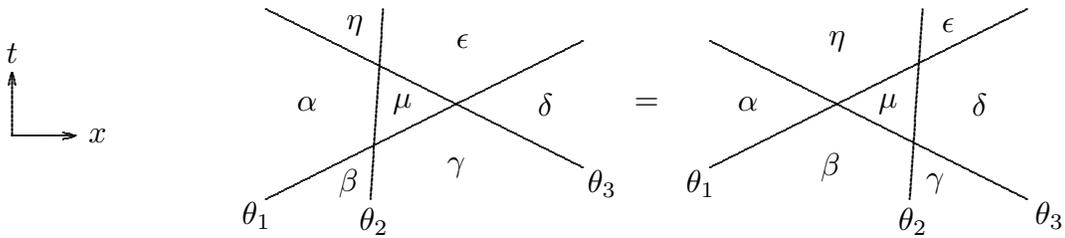
The Yang-Baxter equation~(2.5) (the sum over $\mu$ on both sides of the
equation is suppressed).}
\endinsert

\vskip 0.2cm
Finally, if the kinks in a theory form (stable) bound states, as
indicated by simple poles (with residues of appropriate~\rKar\ sign) of
the scattering amplitudes in the bound state region $\th \in (0,i\pi)$,
then the \sm\ also has to satisfy  {\it bootstrap}
constraints~\rSTW\rTanig. In theories of a single multiplet of kinks this
can of course
happen only if the bound states of the ``constituent'' kinks are again
kinks in the same multiplet (see sect.~4 for an example).
We will treat here only this case; generalizations are straightforward.

\vskip 1.0cm
\midinsert
\centerline{
\beginpicture
\setcoordinatesystem units <1pt,1pt>
\linethickness 4.0pt
\setlinear
\plot 0 30  100 -20 /
\plot 33 -30  37 30 /
\plot 0 -30 100 20 /
\plot 140 30 240 -20 /
\plot 173 -30 174 -13 /
\plot 140 -30 174 -13 /
\plot 174 -13 196 13 /
\plot 196 13  197 30 /
\plot 196 13  230 30 /
\put {$=$} at 120 0
\put {$\alp$} at 13 0
\put {$\beta$} at 26 -25
\put {$\eta$} at 28 25
\put {$\mu$} at 43  0
\put {$\gam$} at 60  -20
\put {$\eps$} at 62 20
\put {$\del$} at 88 0
\put {$\th_1$} at -3 -35
\put {$\th_2$} at 34 -37
\put {$\th_3$} at 106 -25
\put {$\alp$} at 152 0
\put {$\beta$} at 166 -25
\put {$\eta$} at 180 20
\put {$\gam$} at 200  -20
\put {$\eps$} at 205 25
\put {$\del$} at 222 7
\put {$\th_1$} at 137 -35
\put {$\th_2$} at 174 -37
\put {$\th_3$} at 246 -25
\endpicture
}
\vskip 0.6cm
\no  {\bf Fig.~3:}
{\tenpoint
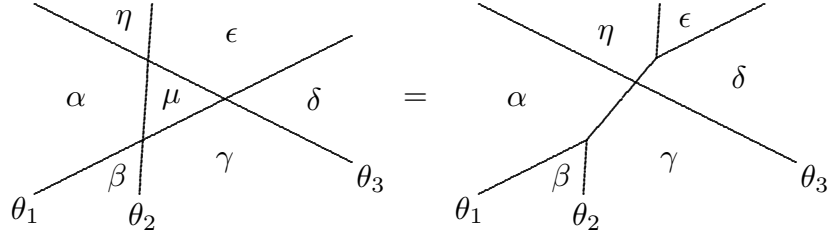
The bootstrap equation~(2.5) 
is obtained by taking the residues
of both sides of the depicted ``equality''
at $\th_1-\th_2=2\pi i/3$ ~(the sum over $\mu$ on the lhs is suppressed).}
\vskip 0.4cm
\endinsert

Let $K_{\alp\gam}$ be a bound state of $K_{\alp\beta}$ and $K_{\beta\gam}$
(in the direct channel).
Considering the 3-kink process $K_{\alp\beta}(\th_1) + K_{\beta\gam}(\th_2) +
K_{\gam\del}(\th_3) \to  K_{\alp\eta}(\th_3) + K_{\eta\eps}(\th_2) +
K_{\eps\del}(\th_1)$ with $\th_1$ and $\th_2$ tuned to the complex
values necessary to  create the intermediate
bound-state kink $K_{\alp\beta}$ on shell (see fig.~3),
one concludes that the following
equations must hold:
\eqn\boot{ \sum_\mu ~g_{\alp\beta\gam}~g_{\alp\mu\gam}~
 S_{\mu\del}^{\gam\eps}(\th+{i\pi\o 3})~S_{\alp\eps}^{\mu\eta}(\th-{i\pi\o 3})
 ~=~ g_{\alp\beta\gam}~g_{\eta\eps\del}~S_{\alp\del}^{\gam\eta}(\th)~~}
(where $g_{\alp\beta\gam}$ can be cancelled on both sides whenever it is
nonzero). Here
 the couplings $g_{\alp\beta\gam}$ are defined (at worst up to signs)
in terms of the residues at the bound-state poles,
\eqn\coupl{ {\rm Res}~S_{\alp\gam}^{\beta\mu}(\th)\Bigl|_{\th=2\pi i/3} ~=~
         i~g_{\alp\beta\gam}~g_{\alp\mu\gam}~.}
Note that~\crossing\ implies that the residue of the crossed-channel pole
is just the opposite of the direct-channel one~\coupl,
\eqn\ccoupl{ {\rm Res}~S^{\alp\gam}_{\beta\mu}(\th)\Bigl|_{\th=\pi i/3} ~=~
         -i~g_{\alp\beta\gam}~g_{\alp\mu\gam}~.}
If all couplings $g_{\alp\beta\gam}$ are real,
the \sm\ is said to be ``1-particle unitary''.
Parity invariance, crossing symmetry,
 and the bootstrap equations imply that the
$(g_{\alp\beta\gam})^2$ are symmetric in $\alp, \beta$ and $\gam$,
{\it as long as} all ~$S_{\alp\gam}^{\beta\beta}(0)=-1$
(cf.~\rourii\ for an analogous statement in diagonal \sm\ theories
of ordinary particles). The latter
presumably holds in all bosonic kink theories, see eq.~(2.10) below.
For fermionic
theories, where $S_{\alp\gam}^{\beta\beta}(0)=+1$, there are some
subtle signs that have to be taken into account,
already in the definition of the couplings~\coupl\ (cf.~\rKar\ for details
in the case of ``ordinary'' scattering theories).

\subsec{The Bethe-Yang Equations}

Consider  a factorizable
\scth\ on a finite space with periodic boundary conditions, in other words
on a cylinder. We are interested in multi-particle energy levels
as a function of the ``volume of the world'', \ie~the circumference $R$ of the
cylinder. In an integrable theory particle number is conserved
and so we can
talk about $N$-particle states for any fixed $N$ --- at least as long as the
notion of particles makes {\it any}
 sense in finite volume, namely when $R\gg 1/m$ and the particles are far
apart. One should think of an
$N$-particle energy eigenstate in large volume as a {\it stationary scattering
state}, \ie\ a superposition of $N$-kink states in our case,
which is invariant
under the scattering of the various kinks. Such a state is characterized by a
set of rapidities $\{\th_1,\ldots,\th_N\}$,
since they are conserved in the scattering processes.
(Note that because of ``mixing'' between different particles of the same mass,
the $\th_i$ cannot be assigned to specific particles).

\medskip
Let $\psi=\psi(x_1,\ldots,x_N)$ be the wavefunction of an $N$-particle state,
and  let  $\psi^{\alp} \equiv\psi^{\alp_1\ldots \alp_N}$  denote
its components with respect
to the basis states  ~$|K_{\alp_1\alp_2}(\th_1) K_{\alp_2\alp_3}(\th_2)\ldots
K_{\alp_N\alp_1}(\th_N)\rangle$.
Thinking of the basis vectors as {\it in}-states we must have
$\th_1>\th_2> \ldots > \th_N$.
The number of $N$-particle states (for fixed $\th_i$) with periodic
boundary conditions will be denoted by $d_N$.

The behaviour of $\psi$ when its arguments
$x_i$ are far apart can be expressed in terms of scattering amplitudes.
Periodic boundary conditions are imposed by,
roughly speaking, stipulating that  the wavefunction
change only by a factor of  $(-1)^F$ when ``carrying a kink
$K_{\alp_k\alp_{k+1}}(\th_k)$ once around the
cylinder''.\foot{Of course, this should not be literally understood as a
physical process, but rather as a mnemonic describing a sequence of
rewritings of the wavefunction.}
(Here $(-1)^F=\pm 1$ for kinks that are bosons or fermions, respectively.)
This leads to the following equations for the allowed $\psi$
\eqn\BATt{ e^{i R m \sinh\th_k} \sum_{\alp:~\alp_k =\beta_{k+1}}
    \tilde{T}_k(\th_1,\ldots,\th_N)_\alp^\beta ~\psi^\alp ~=~
             (-1)^F     ~\psi^\beta ~, ~~~~~~ k=1,\ldots,N, }
where
\eqn\Tt{\tilde{T}_k(\th_1,\ldots,\th_N)_\alp^\beta ~=~
           \prod_{i\neq k}^N
 ~S_{\beta_i \alp_{i+1}}^{\alp_i \beta_{i+1}}(\th_k-\th_i) ~~.}
Here $\alp_{N+1}=\alp_1$ and $\beta_{N+1}=\beta_1$.
(A graphical illustration of a matrix very closely related to $\tilde{T}_k$
will be given below.)
Note that there is no factor corresponding to $i=k$ in~\Tt, since
``$K_{\alp_k\alp_{k+1}}$ does not scatter with itself when taken around the
circle''.
We will refer to~\BATt\ as the {\it Bethe-Yang equations}
(cf.~\rYang\ for analogous equations in a non-relativistic
theory of ``ordinary'' particles).

\medskip
We would like to rewrite~\BATt\ in
a more convenient way, getting rid of the restriction on the sum over the
multi-index $\alp$.
To do so, note that unitarity implies
$\sum_{\del} (S_{\alpha\beta}^{\gam\del}(0))^2 =1$. In all kink theories we
know this is satisfied in the simplest possible way, namely
the \sm\ at zero relative rapidity satisfies
{}~$S_{\alpha\beta}^{\gam\del}(0) =
S_{\alpha\beta}^{\gam\gam}(0)~\del^{\gam\del}
 = \pm \del^{\gam\del}$. ~[In ordinary \sm\ notation this reads
$S_{ab}^{cd}(0)=\pm \del_{ac} \del_{bd}$,
\ie~total reflection
in the low-energy limit.]

Furthermore, as a slight generalization of the arguments and observations
in~\rZamtba\rourii,
we claim that all (non-vanishing)
amplitudes for the scattering of kinks in the same multiplet satisfy
\eqn\allf{    S_{\alp\beta}^{\gam\gam}(0) ~=~ -(-1)^F ~,}
since it amounts to an exclusion principle, \ie\ $\th_i \neq \th_j$ for all
$i \neq j$.  Eq.~\allf\ and the analogous statement in non-kink language is
true in all consistent $(1+1)$-dimensional
QFTs we are aware of, except for free bosons, which are
somewhat singular from various points of view.

\bigskip
We can now rewrite~\BATt\ as
\eqn\BAT{ e^{i R m \sinh\th_k} ~\sum_{\alp}
 ~T(\th_k|\th_1,\ldots,\th_N)_\alp^\beta ~\psi^\alp ~=~
                   - \psi^\beta ~~~~~~~~ k=1,\ldots,N, }
where there is no restriction on the sum over $\alp$. Here $T$ is the
$N$-{\it particle transfer matrix}, the analog of the (inhomogeneous)
row-to-row transfer matrix
(of an IRF model in our kink context)
in statistical mechanics~\rBax,
\eqn\Tdef{ T(\th)_\alp^\beta ~\equiv~ T(\th|\th_1,\ldots,\th_N)_\alp^\beta
 ~=~ \prod_{i=1}^N ~S_{\beta_i \alp_{i+1}}^{\alp_i \beta_{i+1}}(\th-\th_i) ~,}
presented graphically in fig.~4.
\vskip 0.8cm
\midinsert
\vskip 0.5cm
\centerline{
\beginpicture
\setcoordinatesystem units <1pt,1pt>
\linethickness 4.0pt
\setlinear
\arrow <4pt> [.3,.5] from -75 -6 to -63  6
\plot 0 0 180 0 /
\plot 30 30 30 -30 /
\plot 60 30 60 -30 /
\plot 120 30 120 -30 /
\plot 150 30 150 -30 /
{
\put {$t$} at -83 -11
\put {$\beta_1$} at 12 15
\put {$\beta_2$} at 42 15
\put {$\cdot$} at 69 15
\put {$\cdot$} at 81 15
\put {$\cdot$} at 93 15
\put {$\cdot$} at 105 15
\put {$\beta_{N}$} at 133 15
\put {$\beta_1$} at 162 15
\put {$\alp_1$} at 12 -15
\put {$\alp_2$} at 42 -15
\put {$\cdot$} at 69 -15
\put {$\cdot$} at 81 -15
\put {$\cdot$} at 93 -15
\put {$\cdot$} at 105 -15
\put {$\alp_{N}$} at 133 -15
\put {$\alp_1$} at 162 -15
}
\put {$\th$} at -13 0
\put {$\th_1$} at 25 -38
\put {$\th_2$} at 55 -38
\put {$\th_{N-1}$} at 117 -38
\put {$\th_N$} at 146 -38
\endpicture
}
\vskip 0.5cm
\no {\bf Fig.~4:}
{\tenpoint
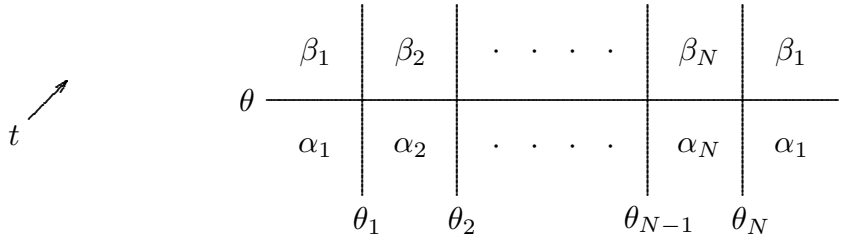
Graphical description of the $N$-kink transfer matrix~(2.10).
Each vertex corresponds
to a 2-kink scattering amplitude, with the time axis
to be thought of as pointing in the NE-direction.}
\vskip 0.2cm
\endinsert

When calculating the finite-volume ground state energy of the theory
using the thermodynamic \BA\ technique, it is the $N\to\infty$ limit of
this transfer matrix that must be considered. Taking
the thermodynamic limit, where also $R\to\infty$ with $N/R$ fixed, certain
simplifications occur which often allow one to determine the form of the
dominant (in the thermodynamic sense) eigenvalues, and derive integral
equations for the distributions of the $\th_i$ minimizing the free energy;
see~\rRSOStba\rFenInt\ for details and examples.

\medskip
Returning to finite $N$, for given $R$ the
\BY\ equations~\BAT~have
solutions only for special $\th_k=\th_k(R)$.
 These equations therefore determine the wavefunctions
$\psi$ and the (total) energies and momenta
\eqn\Ep{E~=~\sum_{k=1}^N ~m \cosh\th_k~,~~~~~ P~=~\sum_{k=1}^N ~m \sinh\th_k}
of finite-volume eigenstates.

A comment on the range of validity of eq.~\Ep\ is necessary. The
Yang-Baxter equations and unitarity imply that for any permutation $Q$
\eqn\TtoN{ \tilde{T}_{Q_1}(\th) \cdot \tilde{T}_{Q_2}(\th) \cdot
 \ldots \cdot \tilde{T}_{Q_N}(\th) ~=~ 1 ~,}
so that eqs.~\BATt\ lead to the quantization of the
momentum        $P$  of a state     as given in~\Ep,
\eqn\pquant{ e^{i P R} ~=~  \bigl( (-1)^F \bigr)^N ~~.}
This equation is in fact exact in {\it any} QFT, irrespective
of the \BY\ equations, because of translational invariance and periodicity.
Therefore eq.~\Ep~for the {\it momentum} is exact.
On the other hand, the above expression for the energy of a state is of course
{\it not} exact for finite $R$
 --- we are neglecting
contributions due to the fact that the behaviour of the wave function $\psi$
can be expressed in terms of scattering amplitudes only when all particles are
far apart, as well as
 tunneling and polarization effects (cf.~sect.~1).
Fortunately, these effects decay
exponentially with $R$ in a massive theory, whereas for a multi-particle state
the leading finite-size corrections to the energy levels determined
by~\BAT\---\Ep\ are ${\cal O}(1/R^2)$, see subsect.~2.4.
All powerlike dependence on $R$ is contained in~\BAT\---\Ep.

\subsec{General properties of $T(\th)$}

The most important property of the \tms\ $T(\th)$, eq.~\Tdef,
is that they commute for different $\th$, with fixed $\th_1,\ldots,\th_N$.
This follows from the Yang-Baxter equations. Any commuting family of \tms,
such as that occurring in~\BAT, therefore has a common set of eigenvectors
$\psi^{(s)}(\th_1,\ldots,\th_N)$ with eigenvalues
$\lam^{(s)}(\th|\th_1,\ldots,\th_N)$, $s$ labelling the different
eigenvectors.\foot{In general it is of course not easy to diagonalize the
transfer matrices
(note that the \sms\ and therefore the transfer matrices of kink theories
are never diagonal).
 In theories with a large global symmetry, in particular non-kink theories,
 this task
simplifies.  Consider, for instance,
the $O(3)$ non-linear $\sigma$-model discussed in~\rLuWo,
where the spectrum consists of an isospin $I$=1 triplet.
In the 9-dimensional space of 2-particle states we can change to a basis
of eigenstates of $I$ and $I_3$,
in which the 2-particle transfer matrix is diagonal.
Furthermore, since the basis change is
accomplished by a matrix with rapidity-independent coefficients, the
eigenvalues (which are 1-, 3- and 5-fold degenerate)
will simply be linear combinations of the 2-particle
\smes. In contrast, for a generic factorizable \sm\ theory
there are not enough global charges to diagonalize
even the 2-particle transfer matrix, and consequently
the change to a diagonal basis will involve
 $\th$-dependent coefficients.}
In the cases we will consider the \tm\ will be meromorphic in all its
arguments (with no poles for real $\th, \th_i$).

Real analyticity and unitarity of $S(\th)$ imply
that the transfer matrix $T(\th)$ is unitary if all $\th-\th_i \in \RR$.
The solutions of~\BAT\ we are interested in correspond to physical states,
so they involve only real $\th_i$. Hence all eigenvalues of $T$ in~\BAT\ are
of magnitude~$1$. This is of course necessary for self-consistency of~\BAT.

Let us denote $\{\th_1,\ldots,\th_N \}$ by $\thb$, and introduce the notation
\eqn\abrev{\eqalign{ T_k(\thb) & ~\equiv~ T(\th_k|\th_1,\ldots,\th_N) \cr
      \lam^{(s)}_k(\thb) & ~\equiv~ \lam^{(s)}(\th_k|\th_1,\ldots,\th_N) \cr
      \del^{(s)}_k(\thb) & ~\equiv~ \del^{(s)}(\th_k|\th_1,\ldots,\th_N)
                  ~=~ -i\ln\lam_k^{(s)}(\thb) \cr }}
for the transfer matrix, its eigenvalues and their phases, respectively.
The $2\pi\ZZ$
ambiguity in the phases has to be fixed at
some, say real, $\thb$ by some convention; everywhere else
the phases are then uniquely determined.

Next note that real analyticty implies
\eqn\Trev{T_k(-\thb) ~=~ T_k(\thb)^\ast   ~~~~~~~~~{\rm for}~~\thb \in \RR^N~.}
Therefore the eigenvalues come in pairs
{}~$\lam_k^{(s)}(\thb), \lam_k^{(\bar{s})}(\thb)$~ related by
 $\thb \to -\thb$ and
complex conjugation, or they are invariant under these operations (we then set
$s=\bar{s}$)
\eqn\lamrev{ \lam_k^{(s)}(-\thb) ~=~  \lam_k^{(\bar{s})}(\thb)^\ast  ~, ~~~
          \del_k^{(s)}(-\thb) ~=~  - \del_k^{(\bar{s})}(\thb)~~
               ({\rm mod}~2\pi )~~~~~{\rm for}~~\thb \in \RR^N. }
Here $s=\bar{s}$ is possible only if ~$\lam_k^{(s)}({\bf 0})  \in \{\pm 1\}$,
obviously.
That the converse is also true follows from Schwarz's reflection
principle.\foot{To prove this we need to know that the eigenvalues
$\lam_k^{(s)}(\thb)$ are real  for at least part of the imaginary
$\thb$ ``axis''. In the cases of physical
interest, where solutions to the \BY\ equations~\BATt\ exist for real $\thb$,
and by analyticity also in a complex neighborhood of $\thb={\bf 0}$,
it is clear
from~\BATt\ that the $\lam_k^{(s)}(\thb)$ have this property.}
The same argument also shows that if we define
\eqn\delt{\tilde{\del}_k^{(s)}(\thb)
  ~\equiv~ \del_k^{(s)}(\thb) -\del^{(s)}(0)~,}
where $\del^{(s)}(0)\equiv\del_k^{(s)}({\bf 0})$, which is clearly
independent of $k$,
then
\eqn\deltprop{\tilde{\del}_k^{(\bar{s})}(\thb) ~=~ \tilde{\del}_k^{(s)}(\thb)
        ~=~ -\tilde{\del}_k^{(s)}(-\thb) ~~~~~~~{\rm for}~~\thb \in \RR^N.}

Let $\psi(\thb)$ be a solution of~\BAT,
$T_k(\thb) \psi(\thb) = \lam_k^{(s)}(\thb) \psi(\thb), ~k=1,\ldots,N$. Complex
conjugation, using~\Trev\---\lamrev, then gives for real $\thb$
\eqn\eveqrev{ T_k(-\thb) \psi(\thb)^\ast ~=~
     \lam_k^{(\bar{s})}(-\thb) \psi(\thb)^\ast    ~, ~~~ k=1,\ldots, N. }
We thus see    explicitly that solutions to the \BY\ equations come in pairs
related by parity, as expected. Namely, if $\thb$ is a solution
so is $-\thb$. These two solutions are distinct if $\thb$ and $-\thb$
are different as unordered sets; they are exactly degenerate in energy
(this will also be true for the exact levels
 when tunneling and off-shell effects are taken into account)
and belong to sectors of opposite total momentum.
This is useful, for instance, in
identifying degenerate and non-degenerate levels
in the zero momentum sector.
To give an example relevant for sect.~4.2, note that
in this sector 3-particle states with rapidities
$\{\th,0,-\th\}$ will not be degenerate, whereas a level
$\{\th_1,\th_2,\th_3 \}$ with all $\th_k \neq 0$ will be degenerate with
the
state with rapidities $\{-\th_3,-\th_2,-\th_1\}$.\foot{There
is another mechanism by which exactly degnerate levels can
arise, namely if the
common eigenspace of all $T_k(\thb)$ is 2- or higher-dimensional
for some eigenvalue $\lam_k^{(s)}(\thb)$
(at least on some sufficiently large submanifold of $\thb \in \RR^N$).
This happens in theories with a sufficiently large      global symmetry,
cf.~footnote~5.}

\medskip

There is one more general property of the transfer matrix that we should
mention. As we will
see in sects.~3--4, it is of some physical interest
to consider what happens if we change the \sm\ as follows:
\eqn\smchange{ S_{\alpha\beta}^{\gam\del}(\th) ~~\to~~
     {\chi_\gam(\th)~\chi_\del(\th) \over \chi_\alp(\th)~\chi_\beta(\th)}~
               S_{\alpha\beta}^{\gam\del}(\th) ~.}
Here the $\chi_\alp(\th)$ are
functions which are, say, meromorphic in $\th$, but otherwise
arbitrary at this point.
  From a lattice model point of view, where the \smes\
correspond to Boltzmann weights of an IRF model,
such a change might be referred to as a
``gauge transformation'', since the partition function with periodic boundary
conditions does not change, obviously. A slightly stronger statement is
that the eigenvalues of the transfer matrix $T(\th)$ do not change
under such a transformation (the eigenvectors do, though).
This is physically interesting since it implies that the
finite-volume multi-kink
spectrum of two scattering theories differing just by factors as in~\smchange\
will be identical, though perhaps only up to terms
that are exponentially small for
large volume.\foot{And using the results of~\rLui\rouriv\
one can check that the same is true for the leading large-volume
finite-size mass corrections
(ignoring tunneling effects, for which we are unable to make such a
statement, cf.~sect.~1).}

\medskip
To explicitly prove that the eigenvalues of $T(\th)$ do not change, note that
under the above transformation
\eqn\Tchange{ T(\th)_\alp^\beta ~~\to ~~\prod_{i=1}^{N}~
     {\chi_{\alp_i}(\th-\th_i)~\chi_{\beta_{i+1}}(\th-\th_i)
    \over \chi_{\beta_i}(\th-\th_i)~\chi_{\alp_{i+1}}(\th-\th_i) }
       ~T(\th)_\alp^\beta ~.}
This implies that the diagonal elements of any power of
$T(\th)$ do not change. In particular, the trace of any power of $T(\th)$
stays the same. Since the traces of the first $n$ powers of an $n\times n$
matrix provide a complete set of invariants,\foot{Except
possibly for a discrete
set of exceptional values of the traces. But in our case, where the
eigenvalues are (branches of)
meromorphic functions, the possibility of
such exceptional values can be ignored.}
we have proved that the eigenvalues of $T(\th)$ do not change.

In physical applications the interest in~\smchange\ is due to the following.
Methods like the quantum-group approach (cf.~sect.~3 for some brief remarks)
allow one to conjecture ``proto-\sms'' for certain QFTs. These objects are
real analytic and satisfy the Yang-Baxter equations, but are not necessarily
unitary or crossing symmetric. Unitarity can often be
restored by multiplying the
``proto-\sm''  by a suitably chosen function of $\th$. The crossing properties
can be changed by factors as in~\smchange. For the transformed \sm\ to still
satisfy the Yang-Baxter equations and be real analytic, the $\chi_\alp(\th)$
have to be of the form $\chi_\alp(\th)=\rho_\alp^{-\th/(2\pi i)}$ with
$\rho_\alp >0$ (up to an $\alp$-independent function of $\th$ which cancels
in~\smchange\ and can therefore be ignored). For $\chi_\alp(\th)$ of this form
the crossing properties are affected as follows
\eqn\modcross{ { S_{\alp\beta}^{\gam\del}(\th) \o
                 S^{\alp\beta}_{\gam\del}(i\pi-\th)}  ~~\to~~
  \sqrt{  {\rho_\alp \rho_\beta \over \rho_\gam \rho_\del} } ~
 { S_{\alp\beta}^{\gam\del}(\th) \o
   S^{\alp\beta}_{\gam\del}(i\pi-\th)} ~.}

Note that for theories in which the kinks are bound states of themselves
the couplings defined in~\coupl\ transform as
{}~$g_{\alp\beta\gam} \to
({\rho_\alp \rho_\gam \o \rho_\beta^2})^{1/6} g_{\alp\beta\gam}$~
under~\smchange, assuming the above form of the
``crossing factors'' $\chi_\alp(\th)$.
In particular, the symmetry of the couplings in
$\alp, \beta, \gam$, and the equality (up to a sign) of the direct and
crossed channel residues of bound-state poles, are valid only if strict
crossing symmetry~\crossing\ holds. On the other hand, one can check that
the bootstrap equations~\boot\ are
{\it not} affected, given the above $\chi_\alp(\th)$.
Examples where the  crossing factors
are important will be encountered in {sects.~3--4.}

\subsec{Solution of the Bethe Ansatz equations}

For any given $\thb=\{\th_1,\ldots,\th_N\}$ the $N$ transfer matrices
$T_k(\thb)$
will have $d_N$ common eigenvectors $\psi^{(s)}(\thb)$ with eigenvalues
$\lam_k^{(s)}(\thb)$ of magnitude~1.
For a given $R$ and eigenvector $\psi^{(s)}(\thb)$
the \BY\ equations restrict the allowed $\th_k$ of a physical state to the
solutions of
\eqn\logBA{ r \sinh\th_k ~+~ \del^{(s)}(\th_k|\th_1,\ldots,\th_N)
    ~=~ 2\pi n_k ~, ~~~~n_k \in \ZZ +{\textstyle \half }~, ~~k=1,\ldots,N,}
which we will refer to as the {\it Bethe Ansatz} (BA) equations, to distinguish
them from the Bethe-Yang equations~\BAT.
Here we introduced the dimensionless ``volume'' of the world
$r=mR$, measured in units
of the Compton wavelength of the kinks.

We assume that for given $s$ and $\{n_k\}$ eqs.~\logBA\ have a unique
solution $\th_k=\th_k(r), ~k=1,\ldots,N$. At least physically this is clear,
by ``continuity of the infinite-volume limit'': As $r\to\infty$ all
$\th_k \to 0$, and the unique solution is $\th_k=(2\pi n_k -\del^{(s)}(0))/r
+{\cal O}(1/r^2)$.

The level of ``type'' $s$ and ``quantum numbers'' $n_k$ will be denoted
by $s(n_1,\ldots,n_N)$.
The allowed $n_1,\ldots,n_k \in \ZZ+{1\o 2}$ are
subject to the constraints that in a sector of total momentum $P=2\pi n/R$
\eqn\constr{ n ~+~  {N\o 2\pi}~ \del^{(s)}(0)
                                    ~=~ \sum_{k=1}^N ~ n_k ~, }
 and $n_1>n_2>\ldots >n_N$,
to comply with the exclusion principle.

It will often be more convenient to label a state not by the quantum numbers
$n_k$, but rather by ~$\tilde{n}_k\equiv n_k - (2\pi)^{-1}\del^{(s)}(0) \in
\ZZ +{1\o 2} - (2\pi)^{-1}\del^{(s)}(0)$.
With these quantum numbers on the rhs of~\logBA, one should
use the phase shifts $\tilde{\del}_k^{(s)}(\thb)$ of~\delt\ on the lhs.
The advantage of the $\tilde{n}_k$ is that now simply
{}~$n = \sum_{k=1}^N \tilde{n}_k$.
We will denote a level $s(n_1,\ldots,n_N)$ equivalently as
$\tilde{s}(\tilde{n}_1,\ldots,\tilde{n}_N)$.

How does one solve eq.~\logBA\ to obtain the $\th_k=\th_k(r)$
characterizing a given level? In general this is of course only possible
numerically.  One can proceed as follows: For given $r$, type $s$, and
allowed $\{n_k\}$, rewrite eqs.~\logBA\ as
\eqn\itnBA{ \th_k ~=~ {\rm arcsinh}\Bigl({2\pi n_k -
         \del^{(s)}(\th_k|\th_1,\ldots,\th_N)
                      \over r} \Bigr) ~~~~~~~ k=1,\ldots,N, }
and then iterate     these equations till they converge to a fixed point. One
subtlety is that when the phases $\del_k^{(s)}(\thb)$
are determined from a numerical
diagonalization of $T_k(\thb)$,  one must make sure not to ``loose track'' of a
 phase when it intersects other phases, or ranges over a region
larger than $2\pi$. For the iteration to converge it is perhaps necessary to
modify the most naive recipe just described on a case by case basis.

\bigskip
A very simple example, that in practice is quite important and can be treated
rather explicitly, is that of 2-kink states
in the sector of zero total momentum.
With ~$n \equiv n_1  \in \ZZ+{1\over 2}$, ~$\th \equiv \th_1$,
{}~$\del(\th) \equiv \del^{(s)}(\th|\th,-\th)$,
we have the following parametric representation
of the 2-kink energy in finite volume
\eqn\iiprE{ (r,E)(\th) ~=~ \Big(~{2\pi n -\del(\th) \over \sinh\th},
                                  ~2m\cosh\th~\Big) ~. }
Solving for $\th=\th(r)$ in a ${1\over r}$ expansion we find
\eqn\iipEexpn{\eqalign{ {E(r)\over m} ~=~
  2  +\left({2\pi\tilde{n} \over r}\right)^2 \Bigl[ 1-{2\del' \over r} +
    {3\del'^2-\pi^2\tilde{n}^2 \over r^2} - & {4\del'^3+(4\pi^2\tilde{n}^2/3)
       (\del'''-4\del') \over r^3} \Bigr] \cr
      & ~~~~~~~~~~~~~~~~~~~ ~+~ {\cal O}({1\over r^6})~\cr } }
where $2\pi\tilde{n}=2\pi n-\del$, and $\del, \del', \del'''$ are the
phase shift and its derivatives evaluated at $\th=0$ (all even derivatives
vanish there).
This expansion can of course easily be extended to
higher orders. Even though this expansion has a finite radius of convergence
around ${1\o r}=0$,
 it is usually not particularly helpful in practice when
comparing with numerical results for the finite-volume spectrum
obtained using a ``small-volume method''
like the TCSA, since the latter typically becomes inaccurate
before the above expansion is useful. Instead, one should employ~\iiprE\
directly, which is correct for all $r$ up to terms exponentially small in $r$;
then there is usually a sizable ``window of overlap'' with the numerical
small-volume methods, as we will see in the next sections.

Note that for {\it arbitrary} levels the first nontrivial term in the large-$r$
expansion is ${\cal O}(r^{-2})$, because of the exclusion principle.
Two different levels  with the same
$\{\tilde{n}_k\}$  will generically be split at  ${\cal O}(r^{-3})$,
which always dominates off-shell  and tunneling contributions.

Finally, we remark that even though energywise the infinite-volume limit is
rather trivial --- any given level  with $N$ particles approaches
$N m$ --- the
corresponding eigenstates, being eigenvectors of $T(0|{\bf 0})$,
remain nontrivial and distinct, generically.
Note that $T(0|{\bf 0})$ is a $d_N \times d_N$ permutation matrix;
its eigenvalues are therefore some collection of $d_N$-th roots of unity.

\newsec{$\phi_{1,3}$-Perturbed Unitary Minimal CFTs}

As for any other perturbed CFTs, we have to specify
the partition function of the unperturbed
CFT and the direction of the perturbation. This is crucial for determining
the nature of the theory, in particular its global symmetry
(which may be spontaneously broken in infinite volume)
and whether the theory is massive or massless.
To simplify the discussion below we will use the following notation:
Let $X_m\pm\phi_{p,q}$ denote the $\phi_{p,q}$-perturbation, in the positive
or negative direction respectively, of the unitary minimal CFT~\rBPZ\
 of central
charge $c_m=1-{6\over m(m+1)}$ and modular invariant partition function
(MIPF)
in the $X=A,D,E$ series~\rModInv.\foot{Since
certain $\phi_{p,q}$ are doubled
in the $D_m$ unperturbed CFTs, one has to specify  more precisely the
perturbation in these cases. The only case this subtlety arises
below is that of $D_5\pm\phi_{1,3}$, by which we will mean the perturbation
by the $\ZZ_2$-even combination of the two $\phi_{1,3}$ fields.
Also, the sign choice for the direction of a perturbation is just
a convention for the sign of the perturbing field.
We use the conventions that are by now standard in the literature.}

Let us first briefly mention the different types of arguments which are
helpful in discovering the kink structure of perturbed CFTs. They give us a
better qualitative understanding of the QFT, and sometimes even provide
crucial hints for constructing its~\sm.

\no (i) Landau-Ginzburg (LG) formulation: It has proven useful in many cases
to describe CFTs by LG potentials involving a small number of ``fundamental''
fields~\rZamLG\rLudCar\rKMS\rNeqtwo.
The perturbed CFT is then obtained by adding to the unperturbed potential
a term corresponding to the LG realization (in the CFT) of the perturbing
field.
In the case of supersymmetry-preserving integrable perturbations of
$N$=2
super-CFTs (for which the potential is believed not to
renormalize), the resulting equations of motion can be treated like classical
field equations and analyzed for kink solutions.
For several models~\rSUSYii\rFenInt\
the results have been used directly to construct
the exact quantum spectrum and \sm.
For theories without $N$=2 SUSY the LG formulation is of more limited use,
since the structure of composite operators and the renormalization properties
of the theories are more complicated.
Nevertheless, in some cases
it does provide intuition about the kink structure~\rZamkink.

\no (ii) Quantum-group motivated restrictions: In
refs.~\rRSG\rEguYan\rSmir\
certain perturbed CFTs are identified as restricted lagrangian
QFTs having $SL_q(2)$ symmetry. The restrictions are made at specific
couplings in the lagrangian which correspond to $q$ being a root of unity.
The kink structure of the restricted theory is then dictated by the tensor
product rules of the (finite number of)
non-singular finite-dimensional highest-weight irreps of the algebra
$U_q[su(2)]$ (see {\it e.g.}~\rAGS).
Namely, the degenerate vacua are labelled by the spins $j$ of
(possibly a subset of) these representations, and a multiplet of kinks
forms a representation of some   spin $s$; kinks in the multiplet interpolate
between the vacuum $j_1$ and vacua $j_2$ appearing in the decomposition of
$j_1 \otimes s$ into $U_q[su(2)]$ irreps.
The quantum group machinary enables one to
go far beyond the determination of the kink structure. Its disadvantage
is that starting from a given perturbed CFT, it is not always straightforward
and easy to construct the restricted QFT which it is to be identified with
(the intuition for choosing the unrestricted theory, coming from the
Feigin-Fuchs construction of the CFT, is not always clear).

\no (iii) Statistical Mechanics: Since perturbed CFTs describe
two-dimensional lattice models in the scaling region around a critical point,
knowledge of the phase
structure of the off-critical lattice models can give useful hints about the
spectrum of the corresponding perturbed CFT.

\vskip 5mm
Returning to the family of $\phi_{1,3}$-perturbed minimal CFTs,
we will now briefly review some of the known facts about the theories
$X_m\pm\phi_{1,3}$.
$A_3\pm\phi_{1,3}$ ($=A_3\pm\phi_{2,1}$) are the two phases of the so-called
Ising field theory (IFT)~\rIFT, which can be obtained by taking the scaling
limit of the Ising model at zero magnetic field from above
or below the critical temperature, respectively.
(From the viewpoint of the underlying lattice model, the two theories
$A_3\pm\phi_{1,3}$ are therefore related by duality.)
$A_3+\phi_{1,3}$ is an interacting
$\ZZ_2$-symmetric theory of a single massive boson
with factorizable \sm\ $S(\th)=-1$.
In $A_3-\phi_{1,3}$ the $\ZZ_2$ symmetry is spontaneously broken,
and the theory is described by
the simplest (factorizable) \sm\ theory of a pair of kinks which interpolate
between two degenerate vacua. Though very simple, it is
instructive to discuss this theory in some detail,
which we will do in subsect.~3.1.

The theories $A_m+\phi_{1,3}$, $m\ge 4$, are
believed~\rZamflow\rLudCar\rKMS\rmlesstba\rourv\
to be massless and flow to the CFTs $A_{m-1}$ in the infrared (IR).
It has recently been argued~\rRav\rourvii\ that similarly $X_m+\phi_{1,3}$
flows to $X_{m-1}$ for $X=D~(m\ge 6)$ and $X=E~(m=12,18,30)$,
while $D_5+\phi_{1,3}$ flows to the CFT $A_4$~\rCarLH88\rourvii.
On the other hand, $X_m-\phi_{1,3}$ is believed to be a massive
theory for any $m\ge 4$ (and any $X$ for which this perturbed CFT is
defined). Factorizable~\sm\ theories of a multiplet of
$2(m-2)$ kinks have been proposed for $A_m-\phi_{1,3}$ independently
in~\rZamkink\ (for $m=4,6$) and~\rRSG\ (for $m\ge 4$).\foot{\sms\
for the theories $X_m-\phi_{1,3}$ with $X\neq A$ have not been explicitly
discussed in the literature, except for $D_6-\phi_{1,3}$ for
which an~\sm\ related through ``orbifolding'' (cf.~\rorbif\ for a
statistical mechanics analogy) to that of
$A_6-\phi_{1,3}$ was constructed~\rZamkink. They
can presumably~\rFenInt\    be    constructed   (for $m$ even)
by ``unitarizing and
crossing-symmetrizing'' the Boltzmann
weights of the critical RSOS lattice models of~\rPasq.}

In~\rRSG\ the scattering theories are constructed by a truncation
of the Hilbert space of the sine-Gordon theory at certain couplings, based on
the quantum group ($SL_q(2)$) symmetry of the model. The restriction leading
to $A_m-\phi_{1,3}$ is obtained when  $q=\exp(i \pi {m+1\o m})$,  and so
the vacua $\alpha$  are labelled by the spins $0,1/2,\ldots,(m/2)-1$
of the $U_q[su(2)]$ highest weight irreps.
A kink carries spin $1/2$, and so only vacua $\alpha$ and $\beta$ such that
$|\alpha-\beta|=1/2$ can be linked by a single kink
$K_{\alpha\beta}(\theta)$.
This situation is graphically described
by the Dynkin diagram of the simple Lie algebra $sl(m)$
where the dots stand for the $m-1$ vacua and the links for the pair of
kinks interpolating between them.
The number of types of $N$-kink states on a circle, $d_N$, can easily be
expressed in terms of the {\it incidence matrix} $I$ of this diagram: Defining
$I_{\alp\beta}=1~(0)$
if the vacua $\alp$ and $\beta$ are (not) linked, we have $d_N=tr(I^N)=
\sum\lam^N$, where $\lam$ are the eigenvalues of $I$.
For the theory $A_m-\phi_{1,3}$, therefore,
$d_N=\sum_{k=1}^{m-1}[2\cos(\pi k/m)]^N$
{}~(note that $d_N=0$ if $N$ is odd, as it should).

The \sm\ conjectured for $A_m-\phi_{1,3}$ is given by~\rRSG\
\eqn\rsgsm{ \eqalign{S&_{\alp\beta}^{\gam\del}(\theta)~=~-{U(\theta)\over\pi i}
   \left({[2\gam+1][2\del+1]\over [2\alp+1][2\beta+1]}\right)^
           {-{\th\over 2\pi i}}  \cr
  & ~ \times \Biggl\{
     \left({[2\gam+1][2\del+1]\over [2\alp+1][2\beta+1]}\right)^{1/2}
   \sinh\left({\theta\over m}\right) ~\del_{\alp\beta}+
     \sinh\left({\pi i-\theta\over m}\right)~\del_{\gam\del} \Biggr\}~
    ,\cr} }
where
\eqn\Udef{ \eqalign{ U(\theta) ~=~& \Gamma\Big({1\over m}\Big) ~
      \Gamma \Big( 1+{i\theta\over \pi m} \Big)
 ~\Gamma \Big(1+{i(\pi i-\theta)\over \pi m}\Big) ~
 \prod_{k=1}^\infty {R_k(\theta)~R_k(\pi i-\theta)\over R_k(0)~R_k(\pi i)}~\cr
  R_k(\theta) ~=~& {\Gamma({2k\over m}+{i\theta\over\pi m})
                      ~\Gamma(1+{2k\over m}+{i\theta\over\pi m}) \over
             \Gamma({2k+1\over m}+{i\theta\over\pi m})
                      ~\Gamma(1+{2k-1\over m}+{i\theta\over\pi m}) }~~,\cr}}
and we use the $q$-number notation
\eqn\qnum{ [a]~=~[a]_q ~=~ {q^a-q^{-a}\over q-q^{-1}}
 ~ =~ (-1)^{a-1}~ {\sin({\pi a\o m}) \over \sin({\pi\o m})} }
(which is unambiguous for integer $a$).
The \sm\ \rsgsm\ is invariant under the $\ZZ_2$-symmetry operation that takes
any vacuum $\alp$ to  $(m/2)-1-\alp$.
Note the ``crossing factors'' in the \sm\ \rsgsm\
(cf.~the discussion in subsect.~2.3)
which ensure strict crossing symmetry~\crossing.

\medskip

Before embarking on a discussion of specific models in the series
$A_m-\phi_{1,3}$, let us point out one feature that is common to the
multi-particle spectrum of all these theories. Namely, as we now show, the
eigenvalues of the $N$-particle transfer matrix always come in pairs of
opposite sign: Note that any  $N$-kink state, written as ~$\alp_1~\alp_2
\ldots \alp_N$~ as a shorthand for $|K_{\alp_1 \alp_2}(\th_1)\ldots
K_{\alp_N \alp_1}(\th_N)\rangle$, has integer (half-integer) $\alp_i$ either
in all even or in all odd positions. In other words, there are two classes of
vacuum configurations.
Let us order our basis states such that the first half have integer $\alp_i$
in the even positions $i=2,4,\ldots$, say, and the second half have integer
$\alp_i$ in the odd positions. Due to the restrictions on neighboring vacua
it is clear that in this basis the transfer matrix is of off-diagonal block
form $\pmatrix{0 & A_1 \cr A_2 & 0\cr}$. This implies that for any
eigenvector of the transfer matrix, written in block form as
$\pmatrix{v_1 \cr v_2\cr}$, there is another one $\pmatrix{v_1\cr -v_2\cr}$
of opposite eigenvalue, as claimed.
When $m$ is odd the  two classes of vacuum configurations
are actually exchanged by the $\ZZ_2$
symmetry of the theory, and so the basis of states can be chosen so that
$v_1=v_2$ are eigenvectors of $A_1=A_2$; each pair of eigenstates consists
therefore of a $\ZZ_2$-even and a $\ZZ_2$-odd state. This is not the case
when $m$ is even, where the two eigenstates in each pair are of the same
parity.

\subsec{Ising Field Theory in the Phase of Spontaneously Broken Symmetry}

This is the case $m=3$ of the above family of theories. There are only
two vacua, labelled by $\alp=0,1/2$, which are exchanged by the
(spontaneously broken in infinite volume) $\ZZ_2$ symmetry.
These vacua correspond to the
all-spins-up and all-spins-down degenerate ground states of the underlying
Ising lattice model. There are only two allowed types of 2-kink states,
$|K_{0,1/2}(\theta_1)K_{1/2,0}(\theta_2)\rangle$ and
$|K_{1/2,0}(\theta_1)K_{0,1/2}(\theta_2)\rangle$, and the two allowed
scattering processes are given by the amplitudes
$S_{~0~,~0~}^{1/2,1/2}(\theta)=S_{1/2,1/2}^{~0~,~0~}(\theta)= -1$.

In general, there are two types of states of any even number $N$ of kinks on
the circle, $|K_{0,1/2}K_{1/2,0}\ldots K_{1/2,0}\rangle$ and
$|K_{1/2,0}K_{0,1/2}\ldots K_{0,1/2}\rangle$,
and none when $N$ is odd. So $d_N=2~(0)$ for $N$ even (odd).
The above two types for $N$ even transform into each other under the $\ZZ_2$
symmetry. Their symmetric and antisymmetric superpositions are
the eigenvectors of the
transfer   matrix $T=\pmatrix{0 & 1\cr 1 & 0\cr}$, with eigenvalues
1 and $-1$ respectively. We therefore have $\del^{(1)}=0$  and
$\del^{(2)}=\pi$. The BA equations~\logBA\ immediately lead       to the
quantization ~$m\sinh\th_k=2\pi \tilde{n}_k/R$~
of the momenta in any allowed multi-kink state, where the $\tilde{n}_k$
are half-odd-integers for the $\ZZ_2$-even
sector (type $s=1$) and integers for the $\ZZ_2$-odd sector ($s=2$).
The corresponding       energies and momenta are therefore simply
\eqn\IFTEPs{ E(R)~=   ~\sum_{k=1}^N
   \sqrt{m^2+\biggl({2\pi \tilde{n}_k \over R}\biggr)^2}~~~~~~,~~~
       P~=~{2\pi\over R} \sum_{k=1}^N \tilde{n}_k~~,}
where $N$ is restricted to be even and the $\tilde{n}_k$ are non-coinciding
integers or half-odd-integers, depending on the sector as above
($P$ is therefore always an integral multiple of $2\pi/R$). In addition to
these ``peculiar'' restrictions, the difference between the energy
levels~\IFTEPs\ and those of a free particle is that the energy
in~\IFTEPs\ is not exact for the Ising field theory (IFT) when off-shell
effects are taken into account.

The Ising field theory in its high- and low-temperature phases (denoted
henceforth as IFT$^{(\pm)}$) is to our knowledge the only
nontrivial massive QFTs for
which the exact energy levels are known. They can be obtained from the
partition function on a cylinder of circumference
$R$ and finite length $L$, with periodic boundary conditions on the
spin field in both directions, which  can be written as~\rFerFi\rouriii\
\eqn\ZIFT{ \eqalign{Z^{(\pm)}(q; r) ~=~ {1\over 2} q^{e_0(r)} ~	\Bigl\{
   & \prod_{n\in\ZZ +1/2} (1+q^{\epsilon_n(r)})~ +
         \prod_{n\in\ZZ +1/2} (1-q^{\epsilon_n(r)})  \cr
  & +~ q^{\hat{e}_1(r)} \Bigl[ \prod_{n\in\ZZ} (1+q^{\epsilon_n(r)}) \mp
           \prod_{n\in\ZZ} (1-q^{\epsilon_n(r)}) \Bigr] \Bigr\} ~~.\cr} }
Here $q=e^{-2\pi L/R}$, $r=mR$ ~($m$ being the mass of the particle/kink in
the $+/-$ case), $\epsilon_n(r) = \sqrt{({r\over 2\pi})^2+n^2}$,
{}~$\hat{e}_1(r) = e_1(r) - e_0(r)$, and
\eqn\ed{  e_{0,1}(r)~ =~  -{r\over 4\pi^2}\int_{-\infty}^\infty
    d\theta ~\cosh\theta~ \ln(1\pm e^{-r\cosh\theta}) }
are the scaling functions corresponding to the ground state and the
first excitation  energies $E_{0,1}(R)=(2\pi/R)e_{0,1}(r)$
in IFT$^{(-)}$ (both decaying exponentially to zero
as $r\to\infty$).

Expanding the expression inside the braces in~\ZIFT\ as
$\sum q^{\hat{e}(r)}$, one can read off all the scaled
energy gaps
\eqn\IFTgaps{ \hat{e}^{(\pm)}(r) = e^{(\pm)}(r)-e_0(r)
                       = \sum_{n\in {\cal N}^{(\pm)}} \epsilon_n(r)
   + \cases{0  & $\ZZ_2$-even sector\cr
            \hat{e}_1(r)~~~& $\ZZ_2$-odd sector~.\cr}     }
Here for the $\ZZ_2$-even sector, which is the same in both phases of
the theory, ${\cal N}^{(\pm)}$ is any set of an even number
($|{\cal N}^{(\pm)}|=0,2,4,\ldots$)
of distinct half-odd-integers. 
For the $\ZZ_2$-odd sector  the ${\cal N}^{(\pm)}$ are sets of
distinct integers with $|{\cal N}^{(+)}|=1,3,5,\ldots$ but
$|{\cal N}^{(-)}|=0,2,4,\ldots$.
Comparing these expressions with~\IFTEPs,
we see that the latter gives the {\it exact} energy gaps in the $\ZZ_2$-even
sector of IFT$^{(-)}$. In the odd sector, on the other the hand,
the results~\IFTEPs\ are not exact. But the exponentially small deviation
$(2\pi/R)\hat{e}_1(mR) =\sqrt{{2m\over \pi R}}~e^{-mR}~[1+{\cal O}((mR)^{-2})]$
is the {\it same} for all levels.
This feature is apparently  special to the IFT.

It is interesting to
find the correspondence between the multi-particle states in
IFT$^{(\pm)}$, specified by the BA quantum numbers
$\tilde{n}_k \in {\cal N}^{(\pm)}$, and the conformal states of the Ising CFT.
To do so, recall~\rCardy\ that the UV limit of the exact scaled energy gaps,
$\hat{e}(0)$, are scaling dimensions in the UV CFT. We therefore
compare~\IFTgaps\ for large $r$, where it agrees with the BA results~\IFTEPs,
with its $r\to 0$ limit. The only ``nontrivial'' fact we need is
$\hat{e}_1(0)=1/8$, the rest follows from $\epsilon_n(0)=|n|$.
Concentrating on the kink phase IFT$^{(-)}$, one concludes that the UV
scaling dimensions $\hat{e}^{(-)}(0)$ of $\ZZ_2$-even states are integers,
thus corresponding to fields in the conformal families of
$[\phi_{1,1}={\sl 1}]$ and $[\phi_{1,3}=\varepsilon]$ (the dimensions
of the ancestor primary fields are
$d_1=0$   
and $d_\varepsilon=1$).
More specifically, $\ZZ_2$-even $N$-kink states in the momentum sector
$P$ ``come from'' $[{\sl 1}]$ ~($[\varepsilon]$) ~if
{}~${N\over 2}+{R\over 2\pi}P$~ is even (odd).  This decomposition of the
$\ZZ_2$-even sector into two subsectors reflects the fact that
$[{\sl 1}]$ ~($[\varepsilon]$) is even (odd) with respect to the
Kramers-Wannier duality.
 On the other hand, the whole $\ZZ_2$-odd
sector originates from $[\phi_{1,2}=\sigma]$ in the UV, as $d_\sigma=1/8$.

The characterization of multi-particle states in terms of their quantum
numbers $\tilde{n}_k$ on the one hand and their UV limits on the other hand,
leads to the following identities for
the level degeneracy in $c={1\over 2}$ unitary highest weight irreps
of the Virasoro algebra:
\eqn\Virdeg{ \eqalign{
 \chi_n^{(0)}~&=~
  (\# {\rm partitions~of}~n~{\rm into~distinct~positive~half~odd~integers})
   \cr
 \chi_n^{(1/2)}~&=~
    (\# {\rm partitions~of}~n+{\textstyle {1\over 2}}~
        {\rm into~distinct~positive~half~odd~integers}) \cr
 \chi_n^{(1/16)}~&=~
    (\# {\rm partitions~of}~n~{\rm into~distinct~positive~integers})~~. \cr} }
Here $n$ is a non-negative integer (we set $\chi_0^{(0)}=\chi_0^{(1/16)}=1$),
and $\chi^{(\Delta)}(q)=q^{\Delta-1/48}\sum_{n=0}^\infty
\chi_n^{(\Delta)} q^n$ ~is the  character of the Virasoro irrep of
highest weight $\Delta$~\rRocha.

What allows one to obtain the exact partition function~\ZIFT,
and consequently~\Virdeg, is the fact that
the IFT can be constructed from the theory of a free massive Majorana fermion
by appropriately summing over the four different boundary conditions for the
fermion field, cf.~eq.~\ZIFT.
The case of other integrable perturbed CFTs is much harder; the exact
(in any analytic form) full finite-volume spectrum is certainly out of reach
by present methods.
However, obtaining the correspondence
between multi-particle states and UV conformal fields, and ultimately
the analog of \Virdeg, does not really require the knowledge
of the exact spectrum at {\it all} volumes. The (numerical) TCSA studies
together with
the large-volume BA analysis described in sect.~2 allow us to find this
correspondence for the low-lying levels, as we will see in the examples
studied in the rest of this paper.

Our results will extend earlier empirical observations~\rLasMar\ for certain
perturbed CFTs described by diagonal scattering theories of ``ordinary''
particles:  It was found that in certain such cases (though not all)
$P$=0 states of two lightest
particles all seem to come from a single conformal family in the UV CFT.
Furthermore, in these cases the BA results rather unexpectedly become exact
in the  $R\to 0$ limit.
In IFT$^{(+)}$ this phenomenon is
explicitly demonstrated by eqs.~\IFTEPs\ and~\IFTgaps: Zero momentum 2-particle
states are in the $\ZZ_2$-even sector, the BA equations
give their exact energies at any $R$, and from the more general comments
above we identify all these states as originating from $[\varepsilon]$ in
the UV.

\subsec{The Subleading Thermal Perturbation of the Tricritical Ising CFT}

We turn now to the theory $A_4-\phi_{1,3}$. Let us label the vacua
by $\alp=-,0,+$ instead of $0,1/2,1$, respectively, as in~\rsgsm.
We also let $\sigma,\sigma'$ take   only   the values $+,-$.
There are eight possible types of 2-kink processes, described by the amplitudes
$S_{00}^{\sigma\sigma'}(\th)$ and $S^{00}_{\sigma\sigma'}(\th)$
which can be written explicitly as
\eqn\Aivsm{ S_{00}^{\sigma\sigma'}(\th)~=~-2^{{\th\over 2\pi i}}
   ~R(\th)~ W^{\sigma\sigma'}(\th) ~~, ~~~~~
         S^{00}_{\sigma\sigma'}(\th)~=~-2^{-{\th\over 2\pi i}}
   ~R(\th)~ W_{\sigma\sigma'}(\th) ~~, }
where
\eqn\AivW{ \eqalign{
    W^{\pm\pm}(\th)~&=~{1\over\sqrt{2}}W_{\pm\pm}(\pi i-\th)~=~
             \cosh{\th\over 4} \cr
    W^{\pm\mp}(\th)~&=~{1\over\sqrt{2}}W_{\pm\mp}(\pi i-\th)~=~
             -i\sinh{\th\over 4}~~, \cr} }
and~\rZamkink\rRSOStba\
(see the first reference in~\rSUSYi\ for the factors ($k-{1\o 2}$))
\eqn\AivR{ \eqalign{ R(\th)~&=~\prod_{k=1}^\infty~(k-{\textstyle {1\over 2}})~
  {\Gamma(k+{i\th\over 2\pi})~\Gamma(k-{1\over 2}-{i\th\over 2\pi}) \over
   \Gamma(k-{i\th\over 2\pi})~\Gamma(k+{1\over 2}+{i\th\over 2\pi})} \cr
      &=~{1\over \sqrt{\cosh(\th/2)}} ~\exp\Biggl\{ {i\over 4} \int_0^\infty
     {dx\over x}~{\sin(x\th/\pi) \over \cosh^2 (x/2)} \Biggr\}~~~.\cr} }
(For explicit computations the alternative form
\eqn\ourAivR{ R(\th)~=~
     {1\over \sqrt{\cosh(\th/2)}}~\exp\Biggl\{ i \int_0^{\th}
           {d\th'\over 2\pi}~{\th'\over \sinh \th'} \Biggr\}~~~ }
is most convenient.) Eq.~\AivR\ is the ``minimal'' solution to the constraints
$R(\th)=R(\pi i-\th)=[\cosh(\th/2)R(-\th)]^{-1}$.

The number of types of $N$-kink states in finite volume with periodic boundary
conditions is 0 if $N$ is odd, and  $d_N=2^{(N/2)+1}$ if $N(>0)$ is even,
of which half are
even and half odd with respect to the $\ZZ_2$ symmetry which exchanges
the $\sigma=+,-$ vacua.

Now consider 2-kink states of zero momentum,
for which we use the basis
\eqn\iipAivbasis{
 \big\{|K_{0{\scsc -}}(\th)K_{{\scsc -}0}(-\th)\ra,\,
       |K_{0{\scsc +}}(\th)K_{{\scsc +}0}(-\th)\ra,\,
       |K_{{\scsc -}0}(\th)K_{0{\scsc -}}(-\th)\ra,\,
       |K_{{\scsc +}0}(\th)K_{0{\scsc +}}(-\th)\ra \big \} }
where $\th>0$.
In this basis the 2-kink transfer matrix appearing in the first of the
two \BY\ equations~\BAT\ is
\eqn\iipAivT{\eqalign{ T(\th| \th,-\th)    ~=~(T(-\th| \th,-\th))^{-1}
     & ~=~
 - \pmatrix{ 0 & 0 & c & 0 \cr
             0 & 0 & 0 & c \cr
             a & b & 0 & 0 \cr
             b & a & 0 & 0 \cr} \cr & ~\equiv~
 -\pmatrix{ 0 & 0 & S^{00}_{{\scsc --}} & 0  \cr
            0 & 0 & 0 & S_{{\scsc ++}}^{00}  \cr
            S^{{\scsc --}}_{00} & S_{00}^{{\scsc +-}} & 0 & 0 \cr
            S_{00}^{{\scsc -+}} & S_{00}^{{\scsc ++}} & 0 & 0 \cr}(2\th) ~~.}}
The $\ZZ_2$-odd eigenvectors are
$(1,-1, \pm\sqrt{(a-b)/c},\mp\sqrt{(a-b)/c})^T$,
while the $\ZZ_2$-even ones are
$(1,1,\pm\sqrt{(a+b)/c},\pm\sqrt{(a+b)/c})^T$. The corresponding
eigenvalues are given by $\pm\sqrt{(a-b)c}$ and $\pm\sqrt{(a+b)c}$,
respectively,
which     explicitly    read
  $\pm R(2\th)\sqrt{\cosh\th}$~ and
$\pm iR(2\th)\sqrt{\cosh\th~f_{1/2}(\th)}$~ where $f_\alp(\th)=
\sinh({\th+i\alp\pi \over 2})/\sinh({\th-i\alp\pi \over 2})$.
As shown in general earlier, the phases
 come in two pairs, with a difference of $\pi$ between the phases in
a pair. Explicitly, one of the phases in the $\ZZ_2$-odd sector is
\eqn\Aivphm{ \del^{(A)}(\th)~=~\del_R(2\th) ~,}
while for the $\ZZ_2$-even sector
\eqn\Aivphp{ \del^{(B)}(\th)~=~\del_R(2\th)-{\textstyle {1\over 2}}
   \arctan(\sinh\th)~~,}
where $\del_R(\th)$,
the phase of $R(\th)$ with the branch choice $\del_R(0)=0$,
can be easily read off from~\ourAivR. These phases are shown in fig.~5.
Note that
$\del^{{(A)}}(\pm\infty) = - \del^{{(B)}}(\pm\infty) = \pm {\pi \over 8}$.

The parametric form of the zero momentum 2-kink energies is then
\eqn\iipAivRE{ (r,E) ~=~ \big(~{2\pi n - \del^{(s)}(\th) \over \sinh\th},
               ~2m\cosh\th~ \big)~,~~~~s\in\{A,B\}, ~~~~n\in
                                        {\textstyle {1\over 2}}\NN ~.}

\no
Remember  that allowing (positive) integer quantum numbers $n$ as well as
half-odd-integers in the above is due to
the fact that the eigenvalues of~\iipAivT\ come in pairs of
opposite sign.

\bigskip
We now briefly discuss 4-particle levels. There are eight ($\theta$-dependent)
4-kink states, so we have four pairs of phases, differing by $\pi$ within
a pair. Allowing the $n_k$ labelling a state to be either all integers or
half-odd-integers, we only have to consider four types of phases, which we will
denote by $s=A,B,C,D$.
Types $A$ and $C$ correspond to $\ZZ_2$-even states whereas $B$ and $D$ to
$\ZZ_2$-odd ones.
We will not present analytic expressions for these phases,
since they are rather
messy. Instead, in fig.~6 we show a typical ``cross section'' of the diagonal
phases. Note that if we use the quantum numbers $\tilde{n}_k$
(cf.~subsect.~2.4), then all $\tilde{n}_k
\in \ZZ$ or $\ZZ+{1\o 2}$ for type $A,B,C$, whereas for type $D$ all
$\tilde{n}_k \in \ZZ+{1\over 4}$ or $\ZZ-{1\over 4}$. In particular, in the
zero momentum sector {\it all}
levels of type $D$ will be doubly degenerate. The
explicit discussion of the lowest 4-particle levels will be presented in the
next subsection where we compare them with numerical results.

\subsec{Comparison with the TCSA}

The truncated conformal space approach~\rYZ\ (TCSA) is a non-perturbative
method to calculate the low-lying energy levels of a perturbed CFT in finite
volume. The idea is to truncate the Hilbert space of the CFT to
a finite-dimensional subspace on which the hamiltonian of the perturbed
theory can be diagonalized numerically. This is to be contrasted with
conformal perturbation theory (CPT), where  the theory is treated
perturbatively in the coupling constant $\lam$,
 but the Hilbert space is not truncated.
The hamiltonian, obtained from
the perturbed CFT action $A_{{\rm CFT}}+\lam\int\Phi$ (where the integration is
over a cylinder of circumference $R$),
simply acts
on the truncated Hilbert space
as an ordinary matrix
\eqn\pCFTH{\langle j |(H_0 +\lam V)(R)|i\rangle = {2\pi \over R}
\Big( (\Delta_i + \bar{\Delta}_i-{c\over 12})~\del_{ij} +
  (2\pi)^{1-y} \lam R^y ~C_{j \Phi i}
   ~\del_{\Delta_i-\bar{\Delta}_i,\Delta_j-\bar{\Delta}_j} \Big)~.}
Here $\{|i\rangle\}$ are
(orthonormalized)
conformal states, created by fields  of left (right) dimension
$\Delta_i~(\bar{\Delta}_i)$,
$y=2-(\Delta_{\Phi}+ \bar{\Delta}_{\Phi})$
is the RG eigenvalue of the perturbing field $\Phi$ which
is assumed to be relevant and spinless, \ie~$y>0$ and $\Delta_\Phi=
\bar{\Delta}_\Phi$, and $C_{j\Phi i}$ are the OPE coefficients~\rBPZ\
of the corresponding fields on the plane.

As the truncation is removed  the eigenvalues $E_i(R)$ of the above matrix
should converge to the exact finite-volume energy levels. There is one
subtlety~\rourv\ though. If $y\leq 1$ the (bare) hamiltonian $H_0 +\lam V$ will
suffer from UV divergences,
so that all of its eigenvalues actually diverge
as the truncation is removed. If one wants to avoid the explicit (and hard)
renormalization of these divergences, one can still calculate energy gaps
$\hat{E}_i(R)=E_i(R)-E_0(R)$ using the TCSA, since the divergences cancel
in the differences of eigenvalues. In view of this fact it is not surprising
that even if $y>1$
energy gaps seem to converge faster than the energy levels
as the truncation is removed. We will therefore always calculate gaps.
Note that $y={4\o 5}$ for $A_4-\phi_{1,3}$, so this is a case where the
energy levels themselves will not converge as the truncation is removed.

The truncation of the Hilbert space is conveniently performed by \eg~ignoring
all states above a certain scaling dimension or level.
Truncation effects become important when considering high levels,
or if the interaction term gets large compared to the unperturbed term
in~\pCFTH. Since the interaction term is proportional
to $\lam R^y$, the TCSA will deteriorate with increasing $R$.

The perturbed theory is necessarily non-scale-invariant,
$|\lam|^{1/y}$ being related
to the mass scale $m$ of the theory. In the cases we will consider,
$m$ will be chosen to be the mass of the kink-multiplet. The coefficient
$\kappa$ relating $\lam$ to $m$, $|\lam| = \kappa m^y$, has been determined
for some perturbed CFTs to high precision by comparison with the
thermodynamic \BA\ (TBA), in
particular $\kappa=|\lam|m^{-4/5}=0.148695516112(3)$
for $A_4 - \phi_{1,3}$~\rRSOStba.  In other cases the mass
gap has to be determined by estimating the $R\to\infty$ limit of levels
corresponding to 1- or 2-particle states.  For more details about the TCSA
and some case studies we refer the reader to~\rYZ\rLMC\rLasMus\rouriv\rourv\
(in particular, TCSA results for $A_4-\phi_{1,3}$ were first obtained
in~\rLMC).

\medskip
In fig.~7 we show the first 35 scaled energy gaps
$\hat{e}_i(r)={R \over 2\pi} \hat{E}_i(R)$, $r=mR$, for $A_4-\phi_{1,3}$
as obtained from the TCSA
in the zero momentum sector, together with our BA results for 2- and 4-kink
levels.
Note that at $r=0$ the TCSA is trivially exact, the $\hat{e}_i(0)$ being
the scaling dimensions $d_i=\Delta_i+\bar{\Delta}_i$ of fields in the UV CFT,
and that the UV fields relevant for the 0-momentum sector are all spinless,
since up to a factor of $2\pi/R$ the momentum of
a state on the cylinder becomes the Lorentz spin of the corresponding field
in the UV limit.
The analytical BA results were obtained as described in subsect.~2.4.

The first two gaps correspond to 0-particle states. In infinite volume
there are three degenerate vacua, but in finite volume they split
exponentially, due to tunneling between the three wells of the potential.
In~\rourv\ we  presented strong
evidence for an {\it exact} expression for the first gap $\hat{e}_1(r)$
in terms of the solution of a TBA-like integral equation
(cf.~also~\rMartexc).
As this is, in particular, an example where the precise form of the large-$r$
behaviour of a 0-particle level is believed to be known, let us mention
that this integral equation predicts
\eqn\eihat{\hat{e}_1(r)~=~{r \o \sqrt{2} \pi^2} K_1(r) ~-~
 {e^{-2r}\o 4\pi^2} \Big(\sqrt{\pi r} -\half -{1\over \pi} +
          {3\o 16}\sqrt{{\pi\over r}} + {\cal O}\big({1\o r}\big)\Big)~.}
Unfortunately, as one sees in the figure, the large-$r$ behaviour is swamped
by truncation errors.

The next few levels are the first of an infinite series of 2-kink states,
which soon begin to overlap with 4-kink states (states of 6 or more
kinks
appear only at larger energies not shown in the figure). The first of
each of the ``doublets'' of 2-kink states correspond to type $A$, the
second to $B$, with the same quantum number $n$; BA results are given for
the $n={1\o 2},1,{3\o 2}, 2,{5\o 2}$ levels. The agreement between
the TCSA and the BA results is quite impressive for the low-lying levels;
for small $r$, large $r$, and higher levels, deviations should be expected
due to the limitations of the two methods. In particular,
although we have not
indicated any estimated errors for the TCSA results in the figure, we have
checked that these errors (as estimated from the variation with truncation
level) can explain all deviations at large $r$. Surprisingly, for some
levels the BA results are actually {\it exact} in the $r\to 0$ limit. We will
come back to this point below.

The first 4-kink level corresponds to
$\At({3\o 2},{1\o 2},-{1 \o 2},-{3\o 2})$, and the analytical results
agree quite spectacularly with the TCSA. For the next 4-kink level,
$\Bt({3\o 2},{1 \o 2},-{1\o 2},-{3\o 2})$, the moderate but noticable
deviations from the TCSA are mainly due to the (in)accuracy of the latter.
We identify the next two TCSA levels shown as corresponding to BA type
$\Ct({3\o 2},{1\o 2},-{1\o 2},-{3\o 2})$, and the degenerate pair
$\Dt({7\o 4},{3\o 4},-{1\o 4},-{9\o 4})$,
$\Dt({9\o 4},{1\o 4},-{3\o 4},-{7\o 4})$,
respectively. That the latter level is exactly degenerate is also clear
from the perturbed CFT point of view, see below. Due to the technical
difficulties involved in keeping track of the phases of type $C$ and $D$
(cf.~the crossings in fig.~6) we have not systematically calculated these
energy levels, but just computed them for a few selected $r$ values to
convince ourselves that the above identification is correct. It is actually
rather clear    from fig.~6 that levels of type $B$ and $C$ of the same
quantum numbers will be almost degenerate, with level $C$ a bit higher,
and that a  level of type $D$ will be quite a bit higher than the former for
large $r$.

The next three 4-kink levels are all of type $A$. The first one corresponds
to $\At(2,1,-1,-2)$, the next one to
$\At({5\o 2},{1\o 2},-{1\o 2},-{5\o 2})$, the last one to the degenerate
pair $\At(2,1,0,-3)$, $\At(3,0,-1,-2)$.
The BA results shown for the first two of these levels
 are somewhat lower than
the TCSA results, for which we blame the latter. (As one can see also for
2-kink levels, the general tendency of the TCSA is to give too high
estimates for the energies as one goes to large $r$ and higher levels).
Above these levels there is a ``combinatorial explosion'' of 4-kink
states; unfortunately the TCSA is not accurate enough anymore to warrant a
further comparison with the BA classification (which
can be continued {\it ad nauseam}).

We should briefly comment on the errors of the TCSA results.
Experience has shown that for large $r$, and in particular when $y<1$,
TCSA results can exhibit certain spurious features, even qualitatively:
In fig.~7, for example, the fact that the second TCSA gap increases for $r>3$
and the crossing of the  first two TCSA 2-particle levels at $r\approx 5.6$
are due to truncation errors. Also, since we generically expect
the TCSA     to become increasingly inaccurate
for higher levels, it seems somewhat surprising how good the first
4-kink level agrees with the BA results in fig.~7; for large $r$, where we
can presumably trust the BA results, the agreement is
 much better than    for        2-particle
levels of comparable energy. We will find a similar feature, even more
pronounced, in the example studied in sect.~4. It is therefore probably
not an accident. Instead, it seems to indicate that the volume at which the
TCSA begins to deteriorate for a given level is not so much determined by
the value of the energy at that volume, but rather
by  the UV scaling dimension of the level.  This is in fact quite plausible,
since the TCSA is non-perturbative in $r$ and the only input required for
this method are CFT data.

In table~1 we compare the UV (CFT) and IR (BA) classification of the first
21 levels in $A_4-\phi_{1,3}$.
For the fields creating the conformal states in the UV limit we use the
following notation, for any perturbed CFT.
 The operators ${\cal L}_{n,i}^{(\Delta)}$,
where $n\in \NN$ and $i=1,\ldots,
\chi_n^{(\Delta)}$ ($=$the degeneracy of level $n$ of the Virasoro irrep of
highest weight $\Delta$ and central charge $c$), 
are certain linear combinations of strings of Virasoro
generators $L_{-n_1}\ldots L_{-n_p}$ with $n_1\ge\ldots\ge n_p\ge 1$ and
$\sum_{j=1}^p n_j=n$. The coefficients in these linear combinations
depend on $n,~\Delta, c$,
and the perturbing field $\Phi$.
They are chosen such that the states
$|i\rangle = {\cal L}_{n,i}^{(\Delta)}|\Delta\rangle$ form a basis of the
irrep of highest weight state $|\Delta\rangle$ at level $n$ in which $\Phi$
is diagonal, namely ~$C_{i_1\Phi i_2}=0$ ~if~ $i_1\neq i_2$.
(The latter requirement is necessary to ensure that the UV states in the table
are the $\lam\to 0$ limit of energy-eigenstates in the perturbed theory.)
The operators $\bar{{\cal L}}_{n,i}^{(\Delta)}$ are defined analogously.

There are two noteworthy features of the results in table~1. The first
is the correspondence between the mechanisms by which exact degeneracies
arise from the conformal and BA points of view. From the UV point of view,
the degeneracy between states
${\cal L}_{n,i_1}^{(\Delta)} \bar{{\cal L}}_{n,i_2}^{(\Delta)}\phi$
and ${\cal L}_{n,i_2}^{(\Delta)} \bar{{\cal L}}_{n,i_1}^{(\Delta)}\phi$
with $i_1\neq i_2$ (where $\phi$ is a spinless primary field of dimensions
$\Delta=\bar{\Delta}$)
will be preserved to all orders of conformal perturbation theory (CPT)
due to parity invariance, and is therefore exact for all $r$
(this is of course also clear from the TCSA point of view).
    From the BA point of view, on the other hand,
levels with quantum numbers
$(\tilde{n}_1,\tilde{n}_2,\ldots,\tilde{n}_N)$ $\neq$
$(-\tilde{n}_N,\ldots,-\tilde{n}_2,-\tilde{n}_1)$ will be degenerate, and
again    because of parity invariance
this degeneracy will be exact for all $r$,
even when the exponential corrections
are taken into account.
We can form parity eigenstates by (anti-)symmetrizing over the two
degenerate BA states, whose UV limits are then identified with
superpositions of definite parity of the corresponding conformal states
(and this is the way in which the UV$\leftrightarrow$IR identification of
degenerate levels should be made in table~1).
Note that from both viewpoints all degeneracies
of this kind are two-fold.
Higher degeneracies are possible if a theory has additional symmetries.

The second feature is the pattern of differences between the BA
$\hat{e}_i(r)$
and the exact ones. We will restrict our remarks mainly to 2-kink levels,
although qualitatively the same picture might be true for states with more
kinks. Because the tricritical Ising CFT $A_4$ has a finite number (six)
of primary fields,  the pattern of UV scaling dimensions $d_i$
repeats itself mod~2 (though with varying degeneracies)
in any sector of fixed spin. Similarly, for 2-kink levels
in a sector of fixed total momentum the BA $\hat{e}_i(0)$ repeat
themselves mod~1 (for generic scattering theories without the pairing of
eigenvalues it would be mod~2). The first four 2-kink levels correspond
to the fields  $\sig', ~\veps', ~L_{-1}\bar{L}_{-1}\sig,
{}~L_{-1}\bar{L}_{-1}\veps$~
in the UV CFT, and the differences
between their scaling dimensions and the corresponding BA $\hat{e}_i(0)$ are
$0,{3\o  40}, {1\o 5}, {3\o 40}$, respectively, cf.~table~1.  This pattern
then repeats itself, presumably forever, with certain descendents in the
same families as the above fields and BA 2-kink levels with higher quantum
numbers.  All other spinless conformal fields
lead to 4- and higher multi-kink levels.

Recalling our discussion (subsect.~3.1) of the BA and the exact energy levels
in $A_3-\phi_{1,3}$, one is naturally led to ask if also here the differences
between the exact and the BA 2-kink levels are just four
``universal'' functions of $r$, in the pattern observed above for $r=0$.
(In particular,
one might wonder if the BA 2-kink levels with UV limit
$\hat{e}_i(0)={7\o 8}$ (mod 2) are in fact exact.)
This is {\it not} the case, though,
as an analysis of eq.~\iipAivRE\ reveals:
Since the large-$\th$ expansion of the phase shifts~\Aivphm --\Aivphp\
involves powers of $e^{-\th}$ multiplied by $\th$, the scaled BA levels have
a small $r$ expansion in powers of $r$ multiplied
by powers of $\ln r$.
Furthermore, the log terms do not cancel in differences
of $\hat{e}_i(r)$ corresponding to BA levels of the same type, differing
only in their quantum numbers.
In contrast, the small-$r$ expansion of the {\it exact} $\hat{e}_i(r)$,
that can be obtained from CPT, is in powers of $r^y=r^{4/5}$, without
any log terms.
[In theories with diagonal \sms,
where the scattering amplitudes are products of
certain universal building blocks (see \eg~\rourii),
 the phases entering the 2-particle BA
equations can be expanded in powers of $e^{-2\th}$. Consequently, the
scaled BA energies have a small-$r$ expansion in powers of $r^2$, which again
contradicts CPT predictions, in general. The Ising field theory
is an exception: The
coincidence of the BA and the exact results for the $\ZZ_2$-even
sector in this case is consistent with the above
remarks because $y=1$ and Kramers-Wannier
duality forces all odd terms in the expansion in $r^y$ to vanish.]

\newsec{$\phi_{2,1}$-Perturbed Unitary Minimal CFTs}

The theories $X_m\pm\phi_{2,1}$ (in the notation of sect.~3)
appear to be massive
for all $m$. In $A_m\pm\phi_{2,1}$ with $m$ even the
perturbation is ``magnetic'', manifestly breaking the $\ZZ_2$ symmetry
of $A_m$, and so the sign of the perturbation is immaterial. For $m$ odd,
on the other hand, the two theories $A_m\pm\phi_{2,1}$ are different, related
to each other by duality. Smirnov proposed~\rSmir\
factorizable $S$-matrices for the theories $A_m-\phi_{2,1}$,
$m\ge 4$ (as well as for the closely related $A_m-\phi_{1,2}$ theories),
based on $SL_q(2)$-restrictions of the imaginary-coupling
$A_2^{(2)}$ affine Toda field theory.
Depending on $m$, the spectrum contains kinks (belonging to
one or more multiplets) and possibly bound
states of kinks (``breathers'').
In the case of $m=4$ the spectrum consists of a single multiplet of
three kinks. For this theory, describing the subleading
magnetic perturbation of the tricritical Ising CFT, Zamolodchikov has
independently proposed~\rslmp\ a scattering theory with the same spectrum
and kink structure
but different amplitudes. One of the purposes of our paper is to decide
which (if any) of the two $S$-matrices for this theory is
correct.\foot{An attempt to resolve this issue has recently been made
in~\rMuss\ by comparing TCSA data for finite-size mass corrections with
 analytical expressions. However, in this case tunneling effects
have to be taken into account (cf.~sect.~1), and in~\rMuss\ an {\it ad hoc}
form of the corresponding contributions was assumed without
justification.
 Therefore  we do not find the analysis of ref.~\rMuss\ satisfactory.}

The CFTs $D_m$ ($m\ge 5$ odd) and $E_m$ ($m=11,17,29$) can also
be perturbed
by $\phi_{2,1}$. Smirnov has shown~\rSmir\ that by an appropriate
``orbifolding'' of the~\sm\ he proposed for $A_5-\phi_{2,1}$,
one can reconstruct
the $\ZZ_3$-symmetric diagonal~\sm\ of~\rKobSwi; this scattering theory
of an ordinary particle and its antiparticle was previously
proposed~\rIJMP\ for the theory $D_5+\phi_{2,1}$.
Except for this case, we are not
aware of any explicit discussion of scattering theories for $X_m\pm\phi_{2,1}$,
$X\neq A$,  in the literature.
In general, we think that ``dual'' relations between various particle
scattering theories and kink theories, and between different kink theories,
deserve further investigation.

\subsec{The $\phi_{2,1}$-Perturbation of the Tricritical Ising CFT
        and its $S$-Matrix}

  From here on we concentrate on the theory $A_4-\phi_{2,1}$. The first
information about the kink structure of this theory came from the
TCSA analysis of~\rLMC. The large-volume spectrum in the zero momentum
sector indicated two degenerate vacua ---
even though there is no $\ZZ_2$ symmetry that relates them ---
 and a single 1-particle
state of exactly half (within the numerical accuracy) the energy of the
2-particle threshold. It was therefore concluded that the spectrum of
the \sm\ theory consists of two kinks, interpolating between the two
vacua, and only one bound state of the kinks (``propagating only in
one of the two vacua'') which is degenerate with them in mass.
Alternatively, it is more instructive to think about the spectrum
as consisting of a single multiplet of {\it three} kinks $K_{01},
K_{10}$, and $K_{11}$, so that the incidence matrix
describing the linking between the two vacua labelled by 0 and 1 is
$I=\pmatrix{0 & 1\cr 1 & 1\cr}$. [This is the natural notation from
the point of view of the $SL_q(2)$ structure of the theory,
as later proposed by Smirnov~\rSmir, where the kinks carry spin 1
and the vacua are characterized by spins 0 and 1 ($q=e^{i\pi/5}$
in our conventions embodied in eq.~\qnum).]

Based on the above spectrum and observations, Zamolodchikov
proposed~\rslmp\ to construct the kink-type \sm\ of the theory using
the Boltzmann weights of the critical hard-hexagon lattice
model~\rBax. The motivation comes from the fact that
 the ``spin'' variables of this IRF model obey exactly
the desired restriction imposed on the kinks (namely the ``spins'' can
be chosen to take the values 0 and 1, and neighboring ``spins'' are not
allowed to be both 0). The nonvanishing Boltzmann weights
corresponding to a face of the square lattice with the ``spins''
$\alp,\gam,\beta,\del$
at its vertices (ordered anti-clockwise starting at the upper left vertex)
are
\eqn\HHBW{  W{\alp~~\del \atopwithdelims[] \gam~~\beta}(u) ~=~
 {\sin({\pi\o 5}  -u)\o \sin{\pi \o 5}}~\del_{\gam\del} ~+~
 {\sin u \o \sin{\pi \o 5}}~\sqrt{{s_\gam s_\del \o s_\alp s_\beta}}~
 \del_{\alp\beta}~, ~~~~~~s_\alp = [2\alp +1]_{q=e^{i\pi/5}} ~, }
where $u$ is the spectral parameter.   Explicitly,  this is
\eqn\HHBWex{\eqalign{
    W{1~~1 \atopwithdelims[] 1~~1}(u)~&=~
         {\sin(u+{2\pi\over 5})\over \sin{2\pi \over 5}} \cr
    W{1~~1 \atopwithdelims[] 0~~1}(u)~&=~W{1~~0 \atopwithdelims[] 1~~1}(u)~=~
        {\sin u\over \sqrt{\sin{\pi \over 5} ~\sin{2\pi \over 5}} } \cr
    W{1~~1 \atopwithdelims[] 1~~0}(u)~&=~W{0~~1 \atopwithdelims[] 1~~1}(u)~=~
        -{\sin(u-{\pi\over 5})\over \sin{\pi \over 5}} \cr
    W{0~~1 \atopwithdelims[] 1~~0}(u)~&=~
         {\sin(u+{\pi\over 5})\over \sin{\pi \over 5}}~~~,~~~~~~
    W{1~~0 \atopwithdelims[] 0~~1}(u)~=~
         -{\sin(u-{2\pi\over 5})\over \sin{2\pi \over 5}} \cr}}
These Boltzmann weights satisfy the Yang-Baxter equation.

Zamolodchikov's proposal is to look for a kink \sm\ of the form
\eqn\zamprop{ S_{\alp\beta}^{\gam\del}(\th)~=~\left({\rho_\gam \rho_\del
        \over \rho_\alp \rho_\beta}\right)^{-{\th\over 2\pi i}}~R(\th)~
    W{\alp~~\del \atopwithdelims[] \gam~~\beta}(\lam \th)~~,}
with $\lam,\rho_\alp$, and $R(\th)$ suitably chosen to satisfy
unitarity~\unit, crossing symmetry~\crossing, and
lead to a direct-channel simple
pole corresponding to one of the kinks at $\th={2\pi i\o 3}$ in all
scattering amplitudes, except $S_{00}^{11}(\th)$ (since there is no kink
$K_{00}$).

Actually, a stronger version of this ``pole constraint'' must hold, namely
the bootstrap equations~\boot\ should be satisfied.
It is convenient to postpone the discussion of the bootstrap,
since already the weaker version of the pole constraint together with
crossing symmetry can be satisfied      only
if $\rho_\alp =s_\alp$
(ignoring an irrelevant $\alp$-independent overall factor),
and ~(i) $R(i\pi-\th)=+R(\th)$ and $\lam\in i({9\o 5}+6\ZZ)$ ~or~
(ii) $R(i\pi-\th)=-R(\th)$ and $\lam\in i(-{6\o 5}+6\ZZ)$.\foot{A similar
ambiguity in the analog of    \Zam's
$\lambda$ shows up in Smirnov's quantum-group
approach. Since he is working with the $A_2^{(2)}$ affine Toda field theory
for arbitrary imaginary coupling, he can fix it as follows: For generic
imaginary coupling the model contains a (lightest) breather, a kink-kink bound
state which is an ``ordinary'' particle. Smirnov demands that its scattering
amplitude with itself be the analytic continuation of that of the  single
boson in the {\it real}-coupling $A_2^{(2)}$ Toda theory of ref.~\rArin.
(It is known that the analogous analytic continuation indeed relates the
scattering amplitude of the single boson in the sinh-Gordon model to that
of the first breather in the sine-Gordon model.)}
Unitarity implies the constraint
\eqn\HHunit{ R(\th)R(-\th)~=~{\sin^2{\pi\over 5}\over
   \sinh({\lam \th\o i}-{\pi i\o 5})~\sinh({\lam\th\o i}+{\pi i\o 5})}~.}

Zamolodchikov chooses $\lam=-{6i\o 5}$.
In this case the constraints on $R(\th)$ have the ``minimal solution''
\eqn\RZ{ R_Z(\th)~=~{-i\sin{\pi\over 5}
                     \over
           \sinh({6\th+\pi i\over 5})}
      ~f_{{4\o 5}}\biggl({12\th \o 5}\biggr) ~~,}
taken  in~\rslmp. Here
 $f_\alp(\th)=\sinh({\th+i\alp\pi\over 2})/\sinh({\th-i\alp\pi\over 2})$.

But               $\lam=-6i/5$ is not the only choice.
Choosing (with hindsight)
 $\lam=9i/5$, the ``simplest'' solution for $R(\th)$ is
\eqn\Rsimp{ R(\th)~=~{\pm i\sin{\pi\over 5} \over
     \sinh({9\th-\pi i\over 5})}~
   f_{-{2\o 5}}\biggl({9\th\over 5}\biggr)
  ~f_{{3\o 5}}\biggl({9\th\over 5}\biggr)~~.}
However,
the resulting \sm\ \zamprop\ exhibits some unwanted simple poles
in the physical strip,
namely at $\th={8i \pi\o 9}$ and $\th={i\pi \o 9}$. There is a cure, though:
 One can always multiply the above $R(\th)$ by
arbitrary ``CDD factors''~\rZams\
of the form  $\prod_\alp F_\alp(\th)$, where
 $F_\alp(\th)=-f_\alp(\th)f_\alp(\pi i-\th)$,  which do not affect the
algebraic
constraints on $R(\th)$.  Muliplying by $F_{-1/9}(\th)$ cancels the unwanted
poles, and a further factor of $F_{2/9}(\th)$ cancels the zeros of
$f_{-2/5}(9\th/5)$ and $f_{3/5}(9\th/5)$ in the physical strip without
introducing new unwanted poles. Choosing the `+' sign in~\Rsimp\ to get the
right signs for the residues of the poles in the full \sm,
cf.~\coupl, we see that all
physical constraints are satisfied by
\eqn\RS{ R_S(\th) ~=~ {i\sin{\pi\over 5} \over
          \sinh({9\th-\pi i\over 5})}~
   f_{-{2\o 5}}\biggl({9\th\over 5}\biggr)~
   f_{{3\o 5}}\biggl({9\th\over 5}~
   \biggr)  F_{-{1\o 9}}(\th) ~F_{{2\o 9}}(\th)~,}
in a ``minimal'' way, in the sense that~\RS\ has the smallest number of poles
{\it and} zeros in the physical strip. And this solution corresponds
to Smirnov's proposal!\foot{In~\rSmir\  $R_S(\th)$  was given in an
integral representation; it was rewritten
in (essentially) the form~\RS\  in ref.~\rMuss.}
Smirnov arrived at his proposal from a completely different
direction~\rSmir, and we --- being aware of his result --- just
rederived it using considerations employed by Zamolodchikov in order
to clarify the relation between the two approaches.

The  function $R_S(\th)$
looks somewhat complicated, in particular since we
had to include the factors $F_{-1/9}(\th)$ and $F_{2/9}(\th)$. The
presence of these factors  becomes less mysterious
when considering the bootstrap equations~\boot. One can check that they
indeed hold for Smirnov's \sm\ (with or without the crossing factors), using
\eqn\Rboot{ R_S(\th+{i\pi\o 3}) ~R_S(\th-{i\pi\o 3}) ~=~
            - {R_S^2(\th) \o 2\cos{\pi \o 5}~ S_{11}^{11}(\th)} ~~.}
Given the factor $F_{-1/9}(\th)$,
necessary to eliminate the unwanted poles,
it is crucial to also include the factor
$F_{2/9}(\th)$ in $R_S(\th)$, otherwise the bootstrap equations are not
satisfied.

There is one more important point we have to
discuss in connection with Smirnov's \sm. In ref.~\rSmir\ there are no
crossing factors, \ie~all $\rho_\alp=1$ in~\zamprop. Smirnov
claims that the resulting violation of crossing symmetry does not make the
theory inconsistent. However, he has not explained how such a violation could
arise in a QFT, which is indeed rather difficult to see.
It would be nice,
to perform a direct check of the presence or absence of the crossing factors.
Unfortunately, as we saw in sect.~2.3, these factors do not affect the
finite-volume spectrum of a QFT (modulo, possibly,     exponentially small
terms), so comparison of BA with TCSA results will not reveal
if they are present or not.

Returning to Zamolodchikov's \sm, we note that for $R_Z(\th)$ eq.~\Rboot\
holds with a {\it plus} sign on the rhs, which leads to a violation of all
the bootstrap equations~\boot\ by a sign.
Since it is just one overall sign, we can restore the
bootstrap equations by reversing the sign of all \smes.
(It would have been a serious problem if some bootstrap equations were
satisfied and others violated by a sign, because
 unitarity, crossing and the Yang-Baxter equations fix all relative
signs.) The overall sign that \Zam\ chose~\rslmp\
 was presumably motivated by the fact
that it leads to positive (imaginary) residues for the direct-channel poles
of \smes\ of the form $S_{\alp\beta}^{\gam\gam}(\th)$.
Such residues correspond
to real couplings $g_{\alp\gam\beta}$ in~\coupl, as expected for a unitary
theory. On the other hand, his sign is such that
$S_{\alp\beta}^{\gam\gam}(0)=+1$ for all allowed $\alp,\beta,\gam$, which
according to~\allf\ means that all kinks are fermions. But then it is hard to
see how they can be bound states of each other.
Changing the overall sign of \Zam's \sm\ solves this problem
(since then all kinks are bosons),
in addition to restoring the bootstrap equations. The \sm\ will however now
violate 1-particle unitarity, \ie~some couplings
$g_{\alp\gam\beta}$ will be purely imaginary.
According to previous experience~\rCMouri\ this suggests that \Zam's \sm\
--- if it describes any consistent theory ---
might be that of a perturbed {\it non}-unitary CFT.

Still,
since the {\it a priori} problems with \Zam's \sm\ are due to signs,
a somewhat subtle issue in \smt\ (cf.~\rKar\rourii\rouriv),
and since problems with one
conjecture do not prove that another (Smirnov's) is correct, it is important
to provide a more direct check of these proposals. This will be done in
the next subsections.

\bigskip

\subsec{Multi-kink states}

Imposing periodic boundary conditions the number of
different types of
$N$-kink states is
$d_N=tr~{0~1 \atopwithdelims() 1~1}^N=({1+\sqrt{5}\over 2})^N +
({1-\sqrt{5}\over 2})^N$, or recursively $d_N=d_{N-1}+d_{N-2}$ with
$d_0=2$ and $d_1=1$.
Note that states with an odd number of kinks are allowed here,
contrary to the theories $A_m-\phi_{1,3}$ of sect.~3.
In particular, there is one type of 1-kink states  $|K_{11}(\th)\rangle$,
with $\th$ quantized in finite volume $R$ according to
$m\sinh\th=2\pi \tilde{n}/R$ with $\tilde{n}\in \ZZ$.
Up to exponentially small
corrections the corresponding energy is just $\sqrt{m^2+(2\pi \tilde{n}/R)^2}$.

\bigskip

For 2-kink states of zero momentum on a circle we use the basis
\eqn\iipHHbasis{
        \big\{~|K_{01}(\th)K_{10}(-\th)\ra,~|K_{10}(\th)K_{01}(-\th)\ra,~
                |K_{11}(\th)K_{11}(-\th)\ra ~\big \}~. }
The notation
\eqn\abcde{ \eqalign{
  a(\th)~&=~\tilde{a}(\th)~=~S_{11}^{11}(\th) \cr
  b(\th)~&=   ~\rho_1^{\th/ (2\pi i)}~
    \tilde{b}(\th)~=~S_{11}^{01}(\th)~=~S_{11}^{10}(\th) \cr
  c(\th)~&=   ~\rho_1^{-\th/(2\pi i)}~
       \tilde{c}(\th)~=~S_{01}^{11}(\th)~=~S_{10}^{11}(\th) \cr
  d(\th)~&= ~\rho_1^{-\th/(\pi i)}~\tilde{d}(\th)~=~S_{00}^{11}(\th)  \cr
  e(\th)~&=  ~\rho_1^{\th/(\pi i)}~\tilde{e}(\th)~=~S_{11}^{00}(\th)
         ~~\cr} }
will be useful below. Here $\rho_1 =2\cos{\pi\o 5}$ for Zamolodchikov's
or the crossing-symmetrized Smirnov  proposal, whereas $\rho_1=1$
for Smirnov's original proposal. Note that the tilded quantities would be the
\smes\ if there where no crossing factors.

The  2-kink transfer matrix now reads
\eqn\iipHHTM{T(\th |\th,-\th) 
      ~=~   - \pmatrix{ 0 & e(2\th) & b(2\th) \cr
                     d(2\th) & 0 & 0       \cr
                                    0 & b(2\th) & a(2\th) \cr} ~~.}
Its three eigenvalues
\eqn\iipHHev{\eqalign{
   \lam(\th) & ~=~ -\dt(2\th) \cr
   \lam_\pm(\th) & ~=~
      \half \Big( \dt-\at \pm \sqrt{(\dt-\at)^2+4\ct\dt}~\Big)(2\th) ~,\cr}}
correspond to the eigenvectors
\eqn\iipHHevec{\eqalign{
  v(\th) &~=~ \Big(\rho_1^{-   \th/(\pi i)}
  ~\rho_1^{   \th/(\pi i)}, ~\rho_1^{    1/2} ~\Big)^T ~, \cr
  v_\pm(\th) &~=~ (1/\sqrt{\chi_\pm}, ~\sqrt{\chi_\pm},
                                 ~\sqrt{\chi_\pm}~z_\pm )(\th)^T~.}}
\no Here
\eqn\chiz{ \chi _{\pm}(\th) ~=~ - {d(2 \th) \over \lam_\pm(\th)} ~~,~~
  ~~~~~    z_\pm(\th) ~=~ - {b(2\th)\over \lam_\pm(\th) + a(2\th)}~~. }
Note that in agreement with our general proof in sect.~2.3 the eigenvalues
are independent of the crossing factors, whereas the eigenvectors are not.

Let us denote the phases of $\lam(\th),~\lam_\pm(\th)$ by~
$\del^{(A)}(\th), \del^{(B)}(\th), \del^{(C)}(\th)$, respectively.
Fig.~8 shows a plot of the three types of phase shifts
for  Smirnov's \sm.
In terms of these phases we have the standard parametric form~\iiprE\ of the
2-kink energy levels.

\bigskip

Let us now turn to 3-kink levels.
For the four-dimensional space of 3-kink states on a circle we use the basis
\eqn\iiipHHbasis{\eqalign{ \big\{
                  &~|K_{01}(\th_1)K_{11}(\th_2)K_{10}(\th_3)\ra,
                   ~|K_{11}(\th_1)K_{10}(\th_2)K_{01}(\th_3)\ra, \cr
                  &~|K_{10}(\th_1)K_{01}(\th_2)K_{11}(\th_3)\ra,
                   ~|K_{11}(\th_1)K_{11}(\th_2)K_{11}(\th_3)\ra ~\big\} ~~.}}
It is possible to give explicit analytic expression for the eigenvalue
phases of the 3-kink \tm\ for general $\th_1,\th_2,\th_3$.
However, except for one of the eigenvalues (see
below) they seem to be rather complicated; we will not write them down, since
in the end the BA equations have to be solved numerically anyhow. Instead,
we show in fig.~9 a representative
``cross section'' of the four phase functions for Smirnov's \sm.

For a special class of 3-kink states the relevant phases have very simple
expressions, as we will presently discuss. For reference, let us
first write down the \tm\ relevant for general 3-kink states:
\eqn\giiipTM{ T(\th|\th_1,\th_2,\th_3) ~=~
   \pmatrix{ 0 & ~c~c~e~ & ~d~b~b~ & ~c~a~b~ \cr
           ~b~b~d~ & 0 & c~e~c & a~b~c \cr   e~c~c & b~d~b & 0 & b~c~a \cr
                                b~a~c & a~c~b & c~b~a & a~a~a \cr} ~~~,}
where the $i$-th factor in each of the triplets of functions is to be
evaluated at $\th-\th_i$.
Note that for $\th \in \{\th_1,\th_2,\th_3\}$ the structure of this matrix
simplifies since $b(0)=0$.

One would expect that in the zero momentum sector
 there are solutions of the \BY\ equations with
$\th_2=0$, and, in fact, that the lowest 3-kink levels are of this form.
This is indeed true.
$\th_2=0$ means that we have to find eigenvectors of ~$T(0|\th,0,-\th)$~
with eigenvalue equal to $-1$, where we have set $\th \equiv \th_1 = -\th_3$.
This eigenvalue
turns out
to have a two-dimensional eigenspace for all $\th$,
 which we can write (projectively)
as ~$(z,1,1,(z c(\th) -e(\th))/b(\th))$~ in terms of a free parameter $z$.
Since transfer matrices $T(\th)$ with different $\th$ can be simultaneously
diagonalized, there must be (at least)
two values of $z$ where this vector is also an
eigenvector of $T(\th|\th,0,-\th)$.
We find that these values are
\eqn\iiipz{z_\pm(\th) ~=~ - {\lam_\pm(\th) \over c(\th) c(2\th)} }
corresponding to the eigenvalues
\eqn\iiipevs{\eqalign{ \lam_+(\th) & ~=~ \ct(\th) \ct(2\th) ~-~
                  {\at(\th)\bt(2\th)\ct(\th) \over \bt(\th) } \cr
             \lam_-(\th) & ~=~ -\ct(\th) \ct(2\th) } }
of  $T(\th|\th,0,-\th)$.
These eigenvalues correspond to the $\th_2=0$ section of the phase shift
functions labelled $A$ and $B$, respectively, in fig.~9.\foot{And this is
perhaps the appropriate moment to remark that for general $\th_i$ the
eigenvalue of type~$B$ of $T_k(\th_1,\th_2,\th_3)$ is
$\lam_k^{{\sc (B)}}(\th_1,\th_2,\th_3)=-\tilde{c}(\th_{ki})
\tilde{c}(\th_{kj})$, where $k,i,j \in \{1,2,3 \}$ are all distinct.}

In terms of the corresponding phases $\tilde{\del}^{(s)}(\th)$, \delt,
the parametric form for the 3-kink energies of type $A,B$ with $\th_2=0$ is
\eqn\iiipHHRE{ (r,E) ~=~ \Big(~{2\pi \tilde{n} - \tilde{\del}^{(s)}(\th)
     \over \sinh\th},
          ~m~(1+2\cosh\th)~ \Big)~,~~~~s\in\{A,B\}, ~~\tilde{n}\in \NN ~.}
All other zero momentum 3-kink levels, for which we do not give explicit
expressions, come in degenerate pairs. They are either the pairs
$\tilde{s}(\tilde{n}_1, \tilde{n}_2, \tilde{n}_3)$ and
$\tilde{s}(-\tilde{n}_3,-\tilde{n}_2,-\tilde{n}_1)$ with $s=A,B$ and
the $\tilde{n}_i$ distinct nonzero integers that sum up to zero, or
$\tilde{C}(\tilde{n}_1, \tilde{n}_2, \tilde{n}_3)$ and
$\tilde{D}(-\tilde{n}_3,-\tilde{n}_2,-\tilde{n}_1)$ with distinct
$\tilde{n}_i \in \ZZ+{1\o 3}$ whose sum vanishes.

\bigskip

\subsec{Comparison with the TCSA}

For $A_4-\phi_{2,1}$ the coefficient $\kappa=|\lam| m^{-9/8}$ has not yet been
determined to high precision by comparing CPT with the TBA.
Performing the TCSA with  $\lam={1\o 2\pi}$ we estimate $m=0.99(1)$
(cf.~\rMuss) by looking at the variation
(with truncation level) of the 1-kink energy for moderate to large volume.
Hence $\kappa = 0.161(2)$. Compared
to the case of $A_4-\phi_{1,3}$ we will see that the greater inaccuracy in
$\kappa$ for $A_4-\phi_{2,1}$ will be more than compensated by the larger value
of $y$, so that the TCSA is accurate up to larger values of $r$.

In fig.~10 we show the first 21  scaled energy gaps in the zero momentum sector
obtained from the TCSA for  the $\phi_{2,1}$-perturbed
tricritical Ising CFT,
together with our BA results for 2- and 3-kinks as
derived from Smirnov's \sm. What is clear right away is that the latter \sm\
receives very strong support from
our results; the \sm\ of~\rslmp\ gives quite
a different picture, definitely not describing $A_4-\phi_{2,1}$.

Since there are two degenerate vacua in infinite volume, there is now only
one gap approaching zero energy exponentially. The next gap corresponds to
the kink $K_{11}$ at zero momentum, which is followed by a series of 2- and
3-kink states, and we can also see the first three 4-kink levels (for
which we have not performed a BA analysis, mainly out of laziness).
The UV and IR classification of levels is shown in table~2.

Because of our detailed discussion of the $A_4-\phi_{1,3}$ case we can be
more brief here. Now the differences between the 2-kink BA $\hat{e}_i(r)$
and the exact gaps at $r=0$ are ${3\o 40}, 0,{3\o 40}$, a pattern which is
repeated mod~2.
So it appears that 2-kink states of type $A,B,C$ correspond to fields
in the families $[\sig'], [\veps'], [\sig]$, respectively, of the tricritical
Ising CFT. One can similarly speculate how
the 3-kink states of different types correspond to specific conformal families:
The 3-kink levels of
types $A$ and $B$ with $\th_2=0$ correspond to the families $[\veps]$ and
$[\veps'']$, respectively, whereas the ``conjugate'' types $C$ and $D$ are
associated to both of the
families $[\sig]$ and $[\veps']$. (It is not so clear, though, what the UV-IR
correspondence is for 3-kink levels of type $A$ and $B$ with $\th_2\neq 0$.)
Again, as in subsect.~3.3, one can show that the differences between the exact
and the BA energy levels can {\it not} be some finite number of ``universal''
functions (corresponding to the types of BA levels, independent of the
$\{n_k\}$), as they are in the Ising field theory of sect.~3.1.

\newsec{Discussion}

The multi-particle finite-volume spectrum provides a very characteristic
``fingerprint''
of a QFT. We discussed how to calculate this spectrum for large volume
in integrable scattering theories of kinks.
For simplicity we considered theories with a particle spectrum consisting of
a single multiplet of kinks, in finite volume with periodic boundary
conditions; these assumptions can easily be generalized.
As comparison with numerical results from the truncated conformal space
approach has shown in several cases, the analytical large-volume
predictions are usually even better than one could have hoped for, in that
they are accurate up to quite small volumes
(in fact, for some levels they become exact again at zero volume).

Applying our results to the case of the subleading magnetic perturbation of
the tricritical Ising CFT, comparison with TCSA data provides very strong
support for the \sm\ conjectured by Smirnov~\rSmir;
the \sm\ conjectured by Zamolodchikov~\rslmp\
definitely does not describe this theory
(as suggested in subsect.~4.1,
it might
 describe a perturbation of a non-unitary CFT). There remains,
however, one subtle question that cannot be answered by this comparison.
Namely, if the original \sm\ of Smirnov should be
multiplied by ``crossing factors'' in order to restore crossing symmetry
in the ``traditional'' form~\crossing.
These crossing factors do not affect the multi-particle spectrum (at least
not the dominant terms). The same is true for 1-particle levels,
and the {\it exact} ground state energy
which can be calculated~\rFendN\
using the thermodynamic Bethe Ansatz
is easily seen not to depend on the crossing factors for {\it any} volume.
Nevertheless, the \sms\
are different and can in fact be distinguished ``experimentally'', by
measuring, for example, time delays (which are proportional to phase shift
derivatives) in scattering experiments.
Since it is hard to see how a non-crossing-symmetric \sm\ could correspond to
a consistent QFT, one
might favor the crossing-symmetrized \sm\
as the correct one. Still, it would be nice to perform some explicit check.

Shifting to more general issues, an important open problem in the study of
(massive) integrable perturbations of CFTs is the relation between the UV
CFT classification of states and the IR Bethe Ansatz classification.
For the examples we have studied, our results show explicitly how this
``UV to IR map'' works for the lowest 20 levels or so
in the zero momentum sector.
In addition to extending this map to arbitrary levels and figuring out the
combinatorics involved
(will this lead to new ways of writing Virasoro characters?), one wonders
if it is possible to understand more conceptually why a given multi-particle
state corresponds to a certain conformal field in the UV. Finally, it remains
to be seen if the simple universal pattern that holds for the differences
between the exact and the Bethe Ansatz finite-volume levels in the Ising
field theory  generalizes to a more sophisticated pattern in other integrable
QFTs.

Throughout the paper we considered only integrable theories.
At least for 2-particle states this restriction is not really necessary.
By increasing the volume
the energy of any level in a generic non-integrable theory will
eventually drop below the threshold for particle creation, with
only 2-particle states scattering elastically among  themselves remaining.
The results of sect.~2 then apply  without any modification.
Of course, for a non-integrable theory there is not much hope of ever
knowing the exact \sm. Therefore one might want to use our results
``in reverse'', namely extract the scattering amplitudes by comparison
with numerical data for the energy levels~\rLuii\rLuWo\rLasMar\
 (and this is just as interesting
for integrable theories whose \sm\ is not known). Looking at 2-particle
levels in the zero momentum sector, one can directly  only obtain the diagonal
phases $\del^{(s)}(\th)=\del^{(s)}(\th|\th,-\th)$. Unfortunately, if one is
just given the energy levels, one does not know how to relate the diagonal
to the asymptotic 2-particle basis of states. Except in rather special cases,
where  the theory is {\it a priori} known to have a large (global) symmetry
so that the relation between these bases is very simple, it will be quite
difficult if not impossible to  extract the physical scattering amplitudes
from the diagonal phases  $\del^{(s)}(\th)$.

\medskip
We mentioned in the introduction that the sine-Gordon (SG) model contains
solitons which are not kinks, in that there are no nontrivial restrictions
on the multi-particle Hilbert space.  We would like to conclude with some
remarks on certain variants of the SG model that {\it do} contain kinks.
Note that to define the (classical)
SG theory we do not only have to specify its lagrangian,
{}~${1\o 2}\partial_\mu\varphi\partial^\mu\varphi-(m_0/\beta)^2
(1-\cos\beta\varphi)$,~ but also the ``target
space'' on which the field $\varphi$ lives. For the standard SG model
$\varphi$ lives on a circle of radius $r'={1\o \beta}$, \ie~one identifies
field configurations differing by a period of the potential.
But what if we let $\varphi$ live
on a circle of radius
$r'={k\o\beta}$, for some  generic $k\in\NN$?\foot{This
possibility was also noticed by Swieca~\rSwi.}
 Let us denote the corresponding
theory by $SG(\beta,k)$. All properties of the ordinary SG model $SG(\beta,1)$
that rely only on local properties of the field $\varphi$, like classical
integrability, will also hold for $SG(\beta,k)$. We expect similar statements
to be true also at the quantum level (provided that
$\beta^2 < 8\pi$). In a semi-classical framework it is clear
that the quantum theory $SG(\beta,k)$ will have a $k$-fold degenerate vacuum.
Labelling the vacua by $\alp=0,{1\o 2},\ldots,{k-1\o 2}$, the theory will have
kinks $K_{\alp\alp'}$ with $|\alp-\alp'|={1\o 2}$ (or ${k-1\o 2}$).  As in
the ordinary SG model there may also be bound states of kinks.

The \sms\ of $SG(\beta,k)$ with different $k$ are also closely related:
Using the obvious $\ZZ_k$ symmetry, as well as time-reversal and
parity invariance, we see that every non-vanishing 2-kink amplitude is equal
to one of the three ($\alp$-independent) amplitudes
$S_{\alp\alp_{{\scsc ++}}}^{\alp_{{\scsc +}}\alp_{{\scsc +}}}(\th)$,
$S_{\alp ~\alp}^{\alp_{{\scsc +}}\alp_{{\scsc +}}}(\th)$,
   $S_{\alp ~\alp}^{\alp_{{\scsc +}}\alp_{{\scsc -}}}(\th)$,
where $\alp_{{\scsc \pm}}=\alp\pm{1\o 2}$, $\alp_{{\scsc ++}}=\alp+1$
(mod~${k-1\o 2}$).
These should be independent of the global properties of the field $\varphi$
and therefore equal to the soliton-soliton, anti-soliton-soliton reflection,
and anti-soliton-soliton transmission amplitudes, respectively, of the
standard SG model~\rZams.
Nevertheless, the global properties of $\varphi$, reflected in the fact that
there are restrictions on multi-particle states, do
give rise to differences between the theories with different $k$, \eg~in their
finite-volume spectra, as is clear by just contemplating
the analysis of sect.~2 for different $k$.

Finally, it is
illuminating to consider the UV limit of the theories $SG(\beta,k)$.
 Changing to standard CFT conventions,
where the massless limit of the (euclidean)
SG field $\sqrt{\pi}\varphi$ is denoted by $X$,
we see that the UV limit of $SG(\beta,k)$ is
the $c=1$ gaussian CFT at radius
$r=\sqrt{\pi} k/\beta$ (in the notation of~\rGinsp).\foot{Note
that all these CFTs can be obtained from the one with
$r=\sqrt{\beta}/\pi$ by ``orbifolding'' with respect to the $\ZZ_k$ subgroup
of the  $U(1)$ (``winding'') symmetry of the gaussian model. By analogy,
$SG(\beta,k)$
can be thought of as a $\ZZ_k$ ``massive orbifold'' of $SG(\beta,1)$.}
Gaussian models at different $r$ differ in their field content,
and considering
$SG(\beta,k)$ as a perturbed CFT, namely the perturbation by the
dimension ${1\o 4}({k\o r})^2={\beta^2\o 4\pi}$ spinless vertex operator
$\cos(kX/r)$,
it is again clear that theories with different $k$ will have different
finite-volume spectra (although the ground state energy is the same).
We plan to provide a more detailed analysis of various aspects of the
theories $SG(\beta,k)$ elsewhere.

\medskip
\bigskip
{\vbox{\centerline{\bf Acknowledgements}}}
\medskip
We would like to thank C.~Ahn, P.~Fendley, V.~Korepin, A.~LeClair and
K.~Schoutens for
discussions. The work of T.R.K.~is supported
by NSERC and the NSF. That of E.M.~by the NSF, grant no.~91-08054.

\bigskip

\vfill\eject
\listrefs

\setbox\strutbox=\hbox{\vrule height15pt depth5pt width0pt}
\centerline{\vbox{\halign{&#\vrule&\strut~#&
               #\vrule&\strut~#&
               #\vrule&\strut~#&
               #\vrule&\strut~#&
               #\vrule\cr\tablerule
& ~~~{\rm UV~field}~$\phi$~~~ && ~~~$d_\phi$~~~
&& ~{\rm BA}~$\tilde{s}(\tilde{n}_1,\ldots,\tilde{n}_N)$~
&&  {\rm BA}~$\hat{e}_i(0)$ & \cr \tablerule\tablerule
& ${{\sl 1}=\phi_{1,1}}$  && 0  && ~~{0-kink~state}~~ &&  & \cr \tablerule
& ${\sig=\phi_{2,2}}$  && ${3\o 40}$ && ~~{0-kink~state}~~ &&  & \cr \tablerule
& ${\veps=\phi_{3,3}}$  && ${1\o 5}$  && ~~{0-kink~state}~~ && & \cr \tablerule
& ${\sig'=\phi_{2,1}}$ && ${7\o 8}$
&& $\At({1\o 2},-{1\o 2})$ && ${1-{1\o 8}}$ &\cr\tablerule
& ${\veps'=\phi_{1,3}}$ && ${6\o 5}$
&& $\Bt({1\o 2},-{1\o 2})$ && ${1+{1\o 8}}$ &\cr\tablerule
& ${L_{-1}\bar{L}_{-1}~\sig}$ && $2+{3\o 40}$ &&
 $\At(1,-1)$ && ${2-{1\o 8}}$ &\cr\tablerule
& ${L_{-1}\bar{L}_{-1}~\veps}$ && $2+{1\o 5}$ &&
 $\Bt(1,-1)$ && ${2+{1\o 8}}$ &\cr\tablerule
& ${L_{-1}\bar{L}_{-1}~\sig'}$ && $2+{7\o 8}$ &&
 $\At({3\o 2},-{3\o 2})$ && ${3-{1\o 8}}$ &\cr\tablerule
& ${\veps''=\phi_{3,1}}$                     && 3  &&
 $\At({3\o 2},{1\o 2},-{1\o 2},-{3\o 2})$ && 3  &\cr\tablerule
& ${L_{-1}\bar{L}_{-1}~\veps'}$ && $2+{6\o 5}$ &&
 $\Bt({3\o 2},-{3\o 2})$     && ${3+{1\o 8}}$ &\cr\tablerule
& ${L_{-2}\bar{L}_{-2}~{\sl 1}}$
       && 4  &&
 $\Ct({3\o 2},{1\o 2},-{1\o 2},-{3\o 2})$ && 4 &\cr\tablerule
& ${\cal L}_{2,1}^{(3/80)} \bar{{\cal L}}_{2,1}^{(3/80)}\sig$
   && ${4+{3\o 40}}$ &&
 $\At(2,-2)$ && ${4-{1\o 8}}$ &\cr\tablerule
& ${\cal L}_{2,2}^{(3/80)} \bar{{\cal L}}_{2,2}^{(3/80)}\sig$
     && ${4+{3\o 40}}$ &&
 $\Bt({3\o 2},{1\o 2},-{1\o 2},-{3\o 2})$ && 4 &\cr\tablerule
& ${\cal L}_{2,1}^{(3/80)} \bar{{\cal L}}_{2,2}^{(3/80)}\sig$
    && ${4+{3\o 40}}$ &&
 $\Dt({7\o 4},{3\o 4},-{1\o 4},-{9\o 4})$ && ${\simeq 4}$ &\cr
& ${\cal L}_{2,2}^{(3/80)} \bar{{\cal L}}_{2,1}^{(3/80)}\sig$
      && ${4+{3\o 40}}$ &&
 $\Dt({9\o 4},{1\o 4},-{3\o 4},-{7\o 4})$ && ${\simeq 4}$ &\cr\tablerule
& ${L_{-2}\bar{L}_{-2}~\veps}$
           && $4+{1\o 5}$ &&
 $\Bt(2,-2)$ && ${4+{1\o 8}}$ &\cr\tablerule
& ${L_{-2}\bar{L}_{-2}~\sig'}$
           && ${4+{7\o 8}}$  &&
 $\At({5\o 2},-{5\o 2})$ && ${5-{1\o 8}}$ &\cr\tablerule
& ${L_{-1}\bar{L}_{-1}~\veps''}$            && 2+3  &&
 $\At(2,1,-1,-2)$                 && 5 &\cr\tablerule
& ${\cal L}_{2,1}^{(3/5)} \bar{{\cal L}}_{2,1}^{(3/5)}\veps'$
              && $4+{6\o 5}$ &&
 $\At({5\o 2},{1\o 2},-{1\o 2},-{5\o 2})$ && 5 &\cr\tablerule
& ${\cal L}_{2,2}^{(3/5)} \bar{{\cal L}}_{2,2}^{(3/5)}\veps'$
              && $4+{6\o 5}$ &&
 $\Bt({5\o 2},-{5\o 2})$ && ${5+{1\o 8}}$ &\cr\tablerule
& ${\cal L}_{2,1}^{(3/5)} \bar{{\cal L}}_{2,2}^{(3/5)}\veps'$
         && $4+{6\o 5}$ &&
 $\At(2,1,0,-3)$ && ${\simeq 5}$ &\cr
& ${\cal L}_{2,2}^{(3/5)} \bar{{\cal L}}_{2,1}^{(3/5)}\veps'$
           && $4+{6\o 5}$ &&
 $\At(3,0,-1,-2)$ && ${\simeq 5}$ &\cr\tablerule}}}
\vskip 3mm
\centerline{{\bf Table~1:}
  UV and IR classification of levels in~$A_4-\phi_{1,3}$.}

\vfill\eject

\setbox\strutbox=\hbox{\vrule height15pt depth5pt width0pt}
\centerline{\vbox{\halign{&#\vrule&\strut~#&
               #\vrule&\strut~#&
               #\vrule&\strut~#&
               #\vrule&\strut~#&
               #\vrule\cr\tablerule
& ~~~{\rm UV~field}~$\phi$~~~ && ~~~$d_\phi$~~~
&& ~{\rm BA}~ $\tilde{s}(\tilde{n}_1,\ldots,\tilde{n}_N)$~
&& {\rm BA}~ $\hat{e}_i(0)$ & \cr \tablerule\tablerule
& ${{\sl 1}=\phi_{1,1}}$  &&  0  && ~~{0-kink~state}~~ &&  & \cr \tablerule
& ${\sig=\phi_{2,2}}$  && ${3\o 40}$ && ~~{0-kink~state}~~ &&  & \cr \tablerule
& ${\veps=\phi_{3,3}}$  && ${1\o 5}$  && ~~{1-kink~state}~~
    &&  & \cr \tablerule
& ${\sig'=\phi_{2,1}}$ && ${7\o 8}$
  && $A({1\o 2},-{1\o 2})$ && ${4\o 5}$ &\cr\tablerule
& ${\veps'=\phi_{1,3}}$ && ${6\o 5}$
  && $B({1\o 2},-{1\o 2})$ && ${6\o 5}$ &\cr\tablerule
& ${L_{-1}\bar{L}_{-1}~\sig}$ && $2+{3\o 40}$ &&
 $\Ct(1,-1)$                && 2 &\cr\tablerule
& ${L_{-1}\bar{L}_{-1}~\veps}$ && $2+{1\o 5}$ &&
 $\At(1,0,-1)$              && 2 &\cr\tablerule
& ${L_{-1}\bar{L}_{-1}~\sig'}$ && $2+{7\o 8}$ &&
  $A({3\o 2},-{3\o 2})$ && ${2+{4\o 5}}$ &\cr\tablerule
& ${\veps''=\phi_{3,1}}$                     && 3  &&
 $\Bt(1,0,-1)$ && ${2+{4\o 5}}$  &\cr\tablerule
& ${L_{-1}\bar{L}_{-1}~\veps'}$ && $2+{6\o 5}$ &&
  $B({3\o 2},-{3\o 2})$     && ${2+{6\o 5}}$ &\cr\tablerule
& ${L_{-2}\bar{L}_{-2}~{\sl 1}}$            && 4  &&
 ~{1st 4-kink~state}~ &&  &\cr\tablerule
& ${\cal L}_{2,1}^{(3/80)} \bar{{\cal L}}_{2,2}^{(3/80)}\sig$
&& $4+{3\o 40}$ &&
 $\Ct({4\o 3},{1\o 3},-{5\o 3})$  && 4 &\cr
& ${\cal L}_{2,2}^{(3/80)} \bar{{\cal L}}_{2,1}^{(3/80)}\sig$
&& $4+{3\o 40}$ &&
 $\Dt({5\o 3},-{1\o 3},-{4\o 3})$ && 4 &\cr\tablerule
& ${\cal L}_{2,1}^{(3/80)} \bar{{\cal L}}_{2,1}^{(3/80)}\sig$
&& $4+{3\o 40}$ &&
 $\Ct(2,-2)$                     && 4 &\cr\tablerule
& ${\cal L}_{2,2}^{(3/80)} \bar{{\cal L}}_{2,2}^{(3/80)}\sig$
&& $4+{3\o 40}$ &&
  ~{2nd 4-kink~state}~              &&   &\cr\tablerule
& ${L_{-2}\bar{L}_{-2}~\veps}$  && $4+{1\o 5}$ &&
 $\At(2,0,-2)$                    && 4 &\cr\tablerule
& ${L_{-2}\bar{L}_{-2}~\sig'}$
&& $4+{7\o 8}$  &&
  $A({5\o 2},-{5\o 2})$           && ${4+{4\o 5}}$ &\cr\tablerule
& ${L_{-1}\bar{L}_{-1}~\veps''}$     && 2+3  &&
 $\Bt(2,0,-2)$                    && ${4+{4\o 5}}$ &\cr\tablerule
& ${\cal L}_{2,1}^{(3/5)} \bar{{\cal L}}_{2,2}^{(3/5)}\veps'$
&& $4+{6\o 5}$ &&
 $\Dt({5\o 3},{2\o 3},-{7\o 3})$  && ${4+{6\o 5}}$ &\cr
& ${\cal L}_{2,2}^{(3/5)} \bar{{\cal L}}_{2,1}^{(3/5)}\veps'$
&& $4+{6\o 5}$ &&
 $\Ct({7\o 3},-{2\o 3},-{5\o 3})$ && ${4+{6\o 5}}$ &\cr\tablerule
& ${\cal L}_{2,1}^{(3/5)} \bar{{\cal L}}_{2,1}^{(3/5)}\veps'$
&& $4+{6\o 5}$ &&
  $B({5\o 2},-{5\o 2})$        && ${4+{6\o 5}}$ &\cr \tablerule
& ${\cal L}_{2,2}^{(3/5)} \bar{{\cal L}}_{2,2}^{(3/5)}\veps'$
&& $4+{6\o 5}$ &&
  ~{3rd 4-kink~state}~ && &\cr\tablerule} } }
\vskip 3mm
\centerline{{\bf Table~2:} UV and IR classification of levels
in~$A_4-\phi_{2,1}$.}

\vfill\eject

\centerline{{\bf Figure Captions}}

\vskip 1cm

\no
{\bf Fig.~5:} Diagonal 2-kink phases $\del^{(s)}(\th)=\del^{(s)}(\th|\th,-\th)$
(in units of $\pi$) for the \sm\ \Aivsm --\AivR,
conjectured for $A_4-\phi_{1,3}$. Shown
are type $A$, eq.~\Aivphm, as the upper line, and type $B$,  eq.~\Aivphp,
as the lower line.

\vskip 5mm
\no
{\bf Fig.~6:} The ``cross section''
$\th \mapsto {1\o \pi} \del^{(s)}(\th|\th,{1\o 3}\th,-{1\o 3}\th,-\th)$ of the
diagonal 4-kink phases of type $s=A,B,C,D$ (labelled from top to bottom for
small $\th$) for the \sm\ \Aivsm\---\AivR.

\vskip 5mm
\no
{\bf Fig.~7:}
Scaled energy gaps $\hat{e}_i(r)$ in the zero momentum sector as obtained
from the TCSA for the $\phi_{1,3}$-perturbed tricritical Ising CFT and the BA
using the \sm\ \Aivsm\---\AivR.
Shown are the first 35 gaps calculated with the TCSA using the 228~states
in the CFT up to level~5 (solid lines), the first ten 2-kink levels from the
BA (dotted lines), as well as BA results for the first  4-kink level of
type $B$ (short dashed line) and  the first
three 4-kink levels of  type $A$ (dashed lines). The next 4-kink level,
for which no BA results are shown, is doubly degenerate.

\vskip 5mm
\no
{\bf Fig.~8:} The three types $s=A,B,C$ (from top to bottom) of diagonal
2-kink phases $\del^{(s)}(\th)=\del^{(s)}(\th|\th,-\th)$ (in units of $\pi$)
for Smirnov's \sm\ conjectured to describe the $\phi_{2,1}$-perturbed
tricritical Ising CFT.

\vskip 5mm
\no
{\bf Fig.~9:} The cross section
$\th \mapsto {1\o \pi}
\del^{(s)}(\th|\th,{1\o 2}\th,
\th_3(\th))$~  where $\sinh\th_3(\th)=
-(\sinh\th+\sinh{1\o 2}\th)$, for
the four types $s=C,D,A,B$ (from top to bottom) of diagonal 3-kink phases
in $A_4-\phi_{2,1}$ (using Smirnov's \sm).

\vskip 5mm
\no
{\bf Fig.~10:}
Scaled energy gaps $\hat{e}_i(r)$ in the zero momentum sector as obtained
from the TCSA for the $\phi_{2,1}$-perturbed tricritical Ising model
and the BA based on Smirnov's \sm.
In addition to the first 21 TCSA gaps calculated  with 228 states
(solid lines), we show the first eight 2-kink levels from the BA (dotted),
and the BA results for the first four 3-kink levels with $\th_2=0$
(dot-dashed), which are alternatingly of type $A$ and $B$, as well as the
first two degenerate pairs of 3-kink levels of type $C$ and $D$ (dashed).
The levels for which no BA results are shown are the first three 4-kink levels.

\bye\end